\documentclass[prx,aps,twocolumn,amsmath,amsmath,amssymb,superscriptaddress,10pt]{revtex4-2}
\usepackage{graphicx}% Include figure files
\usepackage{dcolumn}% Align table columns on decimal point
\usepackage{bm}
\usepackage[usenames]{color}
\usepackage{comment}
\usepackage[unicode=true,
bookmarks=true,bookmarksnumbered=true,bookmarksopen=false,
breaklinks=true,pdfborder={0 0 0},backref=false,colorlinks=true]{hyperref}
\usepackage{physics}
\usepackage{bookmark}
\usepackage{mathtools}
\usepackage{makecell}
\usepackage{subfigure}
\usepackage{multirow}
\usepackage{cleveref}

\hypersetup{
colorlinks=true,
linkcolor=blue,
citecolor=blue,
urlcolor=blue
}

\usepackage{cleveref}
\crefname{appendix}{Appendix}{Appendices}
\crefname{equation}{Eq.}{Eqs.}
\crefname{figure}{Fig.}{Figs.}
\crefname{table}{Table}{Tables}
\crefname{section}{Section}{Sections}
\crefname{mythe}{Theorem}{Theorems}
\crefname{mydef}{Definition}{Definitions}
\newcommand{\dummylabel}[2]{\def\@currentlabel{#2}\label{#1}}
\renewcommand{\paragraph}[1]{\vspace{0.1cm}{\bf \textit{#1}}}

\begin{document}

\title{Symmetry-Driven Floquet Engineering in Multivalley SnS}

\author{Sotirios Fragkos}
\email{sotirios.fragkos@u-bordeaux.fr}
\affiliation{Universit\'e de Bordeaux - CNRS - CEA, CELIA, UMR5107, F33405 Talence, France}

\author{Benshu Fan}
\email{benshu.fan@mpsd.mpg.de}
\affiliation{Max Planck Institute for the Structure and Dynamics of Matter, Center for Free-Electron Laser Science, Luruper Chaussee 149, 22761 Hamburg, Germany}

\author{Umberto De Giovannini}
\affiliation{Max Planck Institute for the Structure and Dynamics of Matter, Center for Free-Electron Laser Science, Luruper Chaussee 149, 22761 Hamburg, Germany}
\affiliation{Università degli Studi di Palermo, Dipartimento di Fisica e Chimica -- Emilio Segrè, via Archirafi 36, I-90123 Palermo, Italy.}

\author{Dominique Descamps}
\affiliation{Universit\'e de Bordeaux - CNRS - CEA, CELIA, UMR5107, F33405 Talence, France}

\author{Stéphane Petit}
\affiliation{Universit\'e de Bordeaux - CNRS - CEA, CELIA, UMR5107, F33405 Talence, France}

\author{Hannes Hübener}
\affiliation{Max Planck Institute for the Structure and Dynamics of Matter, Center for Free-Electron Laser Science, Luruper Chaussee 149, 22761 Hamburg, Germany}

\author{Angel Rubio}
\email{angel.rubio@mpsd.mpg.de}
\affiliation{Max Planck Institute for the Structure and Dynamics of Matter, Center for Free-Electron Laser Science, Luruper Chaussee 149, 22761 Hamburg, Germany}
\affiliation{Nano-Bio Spectroscopy Group, Departamento de Fisica de Materiales, Universidad del País Vasco UPV/EHU, 20018, San Sebastián, Spain}
\affiliation{Initiative for Computational Catalysis, The Flatiron Institute, Simons Foundation, New York City, NY 10010, United States of America}

\author{Samuel Beaulieu}
\email{samuel.beaulieu@u-bordeaux.fr}
\affiliation{Universit\'e de Bordeaux - CNRS - CEA, CELIA, UMR5107, F33405 Talence, France}

%\begin{document}

\begin{abstract}
Coherent interactions between time-periodic electromagnetic fields and materials offer a powerful platform for engineering light-matter hybrid Floquet states with tailored functionalities. In particular, the ability to manipulate the wavefunction symmetry of such Floquet states has recently emerged as a new frontier in the field of nonequilibrium control of quantum materials. Here, we investigate symmetry-driven Floquet engineering in bulk multivalley semiconductor tin sulfide (SnS) using time-, polarization-, and angle-resolved extreme ultraviolet photoemission spectroscopy, group-theory analysis, and time-dependent density functional theory. We demonstrate that the material's inherent symmetry gives rise to pronounced symmetry-driven photoemission selection rules for both equilibrium bands and light-induced Floquet states, which we probed through nonequilibrium linear dichroism in extreme ultraviolet photoemission. By leveraging the interplay between crystal and driving-light symmetry, we establish deterministic control over the parity of Floquet--Bloch states. Indeed, we show that the symmetry of Floquet--Bloch states can be fully controlled by the relative alignment between the drive polarization and the crystal axes, enabling selective parity inversion with respect to the equilibrium valence and conduction bands. Furthermore, we show that symmetry-driven parity engineering allows for polarization- and valley-selective band renormalization. These findings advance the understanding of guiding principles for wavefunction symmetry engineering, providing pathways for selectively controlling both the parity and renormalization of electronic states in quantum materials using tailored electromagnetic fields.

\end{abstract}
\date{\today}

\maketitle

\section{Introduction}
The use of light to control quantum materials has become an increasingly active area of research, offering new avenues for realizing and manipulating emergent states of matter~\cite{Basov17, Torre21, Bao2022}. A prominent nonequilibrium control paradigm is Floquet engineering, which exploits the coherent interaction between matter and time-periodic electromagnetic fields to tailor electronic states beyond equilibrium.~\cite{Oka09, Kitagawa11, Lindner11, Wang13, Sie15, Mahmood16, Oka19, Rudner20, Reutzel20, Shan21, McIver20, Aeschlimann21, Zhou23, Zhou23_2, Ito23, Kobayashi23, liu23, weitz24, Bao2024, fan2024chiral}. By appropriately choosing the amplitude, frequency, and polarization of the drive, such periodic perturbations can be used to induce light-dressed quantum states with tailored properties~\cite{Claassen16, Hubener17, Lui18, Schuler22, Trevisan22, Strobel23, Zhou23, Zhou23_2, ValleyFloquet24}. Time- and angle-resolved photoemission spectroscopy (trARPES) has become an indispensable tool for exploring these driven quantum systems, offering direct access to band replicas, light-induced hybridization gap opening, and band renormalizations~\cite{Wang13, Ito23, Zhou23, Zhou23_2, Bao2024-le, Bao2024-lq}. Moreover, while Floquet states result from the periodic coherent dressing of Bloch electrons in the crystal, outgoing photoelectrons can also be dressed by the pump field, giving rise to so-called Volkov states. Despite their distinct microscopic origins, both transitions end up at the same final energy and momentum, resulting in quantum path interferences that manifest as a pump-polarization-dependent modulation of the photoelectron angular distributions -- a hallmark of Floquet--Bloch physics in trARPES experiments~\cite{Park14, Mahmood16, choi24, merboldt24, ValleyFloquet24, Bao2025-zb}.

Beyond the search for conventional signatures of Floquet–Bloch state formation in photoemission spectra, an emerging frontier in the field focuses on controlling the wavefunction symmetry of light-dressed states. Indeed, recent studies on black phosphorus have demonstrated that the symmetry of Floquet states can be dynamically reconfigured by tailoring the polarization of the driving light field~\cite{Zhou23,Bao2024,fan2025floquet}. In these systems, the photon-dressed sidebands inherit symmetry properties dictated by both the underlying lattice and the pump laser geometry, enabling selective switching of parity and pseudospin relative to equilibrium bands~\cite{Zhou23}. For instance, in thin-film black phosphorus, the application of linearly polarized mid-infrared pulses along specific crystallographic directions leads to Floquet sidebands with opposite parity to the original valence band (VB), as directly visualized via trARPES~\cite{Bao2024}. This parity switching, mediated by the interplay between lattice symmetry and light polarization, allows for momentum-selective enhancement or suppression of sideband spectral weights, forming distinct “hot spot” features near the Brillouin zone (BZ) center. Such phenomena are captured by symmetry-driven Floquet optical selection rules, which provide a general framework for interpreting and predicting light-induced spectral distributions~\cite{fan2025floquet}. In a manner analogous to observations in black phosphorus, a study in 2H-WSe$_2$ upon below-bandgap Floquet engineering with circularly polarized driving pulses revealed a distinct modification of the orbital character within the Floquet sideband~\cite{ValleyFloquet24}. Such orbital mixing has been shown to arise directly from coherent light–matter dressing under below-bandgap driving conditions, highlighting the role of optical selection rules and quantum geometry in shaping Floquet states~\cite{ValleyFloquet24}. However, current implementations of symmetry-driven Floquet engineering have been largely restricted to states near the BZ center~\cite{fan2025floquet, Bao2024}, leaving open the question of how symmetry selection rules manifest in anisotropic materials across the full momentum space. In multivalley materials, where inequivalent valleys host distinct symmetries, the interplay between crystal symmetry and light polarization can generate valley-dependent selection rules that remain largely unexplored. A general experimental and theoretical framework capable of resolving and controlling the parity of light-dressed states across the full BZ is therefore still missing.

In this work, we investigate symmetry-guided Floquet engineering in bulk multivalley tin sulfide (SnS), an anisotropic layered semiconductor with low crystal symmetry, using time-, polarization-, and angle-resolved extreme ultraviolet (XUV) photoemission spectroscopy combined with group-theory analysis and time-dependent density functional theory (TDDFT)~\cite{Runge84, andrade2015real, tancogne2020octopus, fan2025floquet}. By exploiting symmetry-driven selection rules of light-matter interactions, we demonstrate deterministic control of Floquet–Bloch states' parity (defined as the eigenvalue of mirror symmetry) at high-momentum valleys. Using XUV photoemission linear dichroism across the entire BZ, we show that Floquet sidebands can acquire a parity opposite to that of the valence and conduction bands, establishing that the wavefunction parity of Floquet states is governed by the combined symmetry of the crystal structure and the time-periodic driving field. In addition, we observe polarization- and valley-dependent Floquet-induced band renormalization under below-gap excitation. These findings advance the understanding of general symmetry-based principles for inducing, controlling, and probing nonequilibrium wavefunction symmetries using light-matter interactions.

\section{Results}

\subsection{Experimental and theoretical methodologies}

\begin{figure*}[!t]
\centering
\includegraphics[width=18cm]{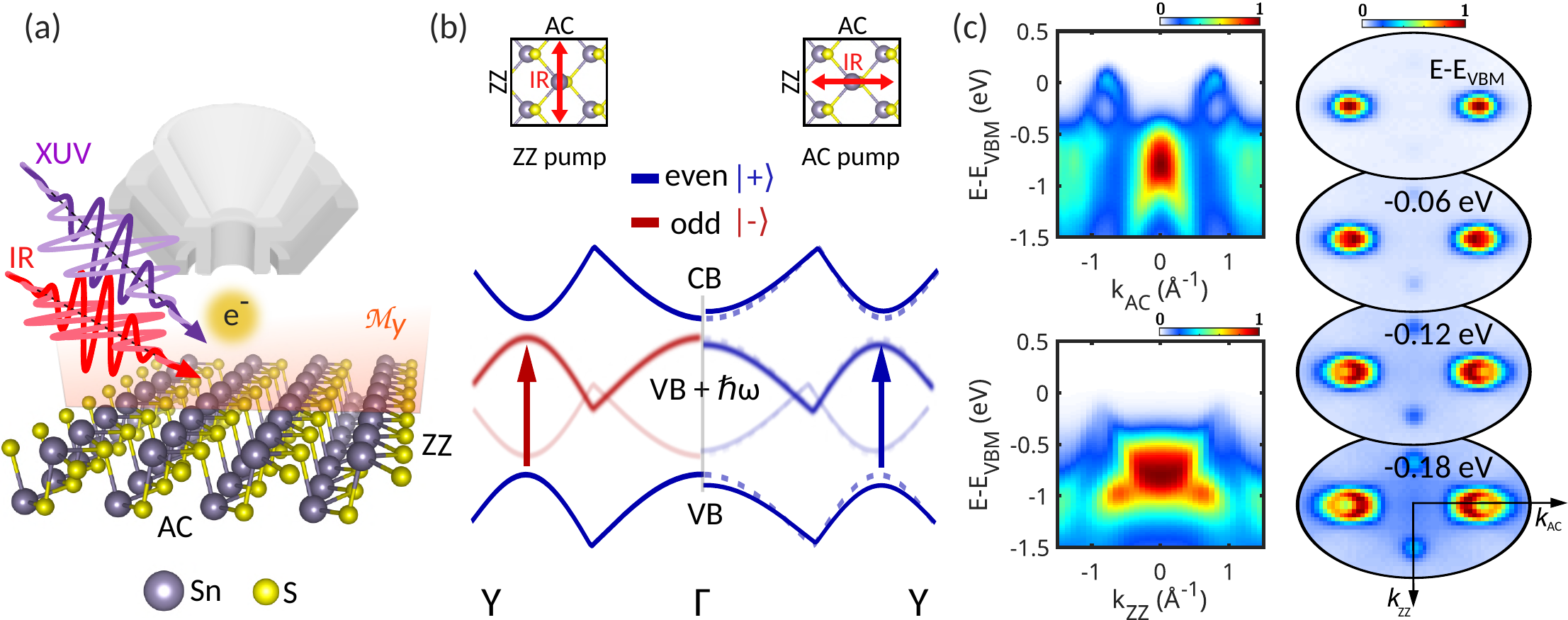}
\caption{\textbf{Experimental Setup and Static Band Mapping of SnS}. \textbf{(a)} Polarization-tunable infrared pump (1.2~eV, 135~fs, 1.3 mJ/cm$^2$) and polarization-tunable XUV (21.6~eV) probe pulses are focused onto a bulk SnS sample, in the interaction chamber of a time-of-flight momentum microscope, at an incidence angle of 65$^{\circ}$, with the light incidence plane along the crystal mirror plane $\mathcal{M}_{y}$ (AC direction). \textbf{(b)} Concept of symmetry-driven Floquet engineering in multivalley SnS, where the parity of Floquet--Bloch states, and band renormalization can be controlled by the relative alignment of the drive polarization and the crystal axes. The square insets depict the real-space in-plane structure of SnS with pump polarization (red arrows) along ZZ or AC directions. \textbf{(c)} Static energy-momentum cuts measured along AC and ZZ directions, integrated over all XUV polarization angles, and associated constant energy contours (CEC) at different binding energies, highlighting the multivalley nature of SnS.}
\label{Fig1}
\end{figure*}

Figure~\ref{Fig1}(a) shows a schematic of the experimental setup. Polarization-tunable infrared (IR, 1.2~eV, 135~fs, $\sim$1.3~mJ/cm$^2$) pump and polarization-tunable XUV (21.6~eV) femtosecond probe pulses are focused onto a bulk SnS sample at room temperature in the interaction chamber of a time-of-flight momentum microscope. The incidence angle on the sample is 65$^{\circ}$ and the light incidence plane is aligned with the crystal mirror plane ($\mathcal{M}_y$) along the armchair (AC) direction. To perform dichroic ARPES measurements, we continuously rotate a visible (515 nm) half-wave plate (HWP) placed upstream of the high-order harmonic generation chamber while recording the photoemission intensity in real time. The linear polarization axis of the green driving pulses is transferred to the XUV high-order harmonics. Each detected electron is tagged with the instantaneous angular position of the HWP. To extract the full polarization-dependent information contained in the photoemission yield from different states through linear dichroism ARPES (LD-ARPES), we perform a Fourier analysis of the periodically modulated signal, following the procedure described in Ref.~\cite{Fragkos2025-ob}.

Bulk SnS is a member of the group-IV tin family (SnX, X=S, Se), which crystallizes in the orthorhombic \textit{Pnma} space group with puckered layers analogous to those in black phosphorus, giving rise to pronounced anisotropy between the AC and zigzag (ZZ) directions and to the emergence of multiple addressable valleys~\cite{Rodin2016-ie, Hanakata2016-bi, Lin2018-do, Chen2018-wv, Thanh_Tien2024-oo, Tolloczko2025-kn, Hashmi2025-rf, Pan2025-jx}. Owing to this layered geometry, the fundamental symmetry properties of the low-energy electronic states are largely preserved when transitioning from bulk to monolayer. In monolayer SnS, the valence band maxima (VBM) at the $\Gamma$ point and the Y$^\prime$ valley are nearly degenerate, whereas in the bulk the VB maximum resides at the Y$^\prime$ valley. Importantly, the corresponding VB wavefunctions along the $\Gamma$–Y line exhibit the same parity with respect to the mirror plane $\mathcal{M}_y$ in both systems. This symmetry equivalence thereby justifies the use of monolayer SnS as a faithful minimal model for the TDDFT simulations, while capturing the symmetry-driven Floquet physics observed experimentally in the bulk [see S1 and S2 of the Supplementary Materials (SM)~\cite{SM} for more symmetry analysis]. Figure~\ref{Fig1}(b) introduces the concept of symmetry-driven Floquet engineering in SnS. As will be discussed in detail below, controlling the IR polarization, and thus the parity of the driving field with respect to specific crystal symmetry operations, enables selective engineering of the symmetry of Floquet sidebands as well as a polarization- and valley-selective band renormalization. Figure~\ref{Fig1}(c) shows the static electronic band structure measured along the $k_{\rm AC}$ and $k_{\rm ZZ}$ directions, integrated over all XUV polarization angles, as well as the corresponding constant energy contours (CECs) at different binding energies. These measurements reveal two pairs of non-degenerate VBM along the $k_{\rm AC}$ and $k_{\rm ZZ}$ directions, which we refer to as the Y$^\prime$ and X$^\prime$ valleys, respectively. It should be mentioned that the photoemission signals, both experimental and theoretical, presented in this work, are symmetrized to enhance statistics and remove spurious symmetry-breaking contributions caused by non-normal incidence. This symmetrization does not affect the inferred parity assignment of the electronic states.

To elucidate the symmetry of both equilibrium electronic bands and nonequilibrium Floquet(-Volkov) states in SnS, we performed time- and polarization-dependent XUV ARPES measurements supported by associated state-of-the-art TDDFT calculations (see methods section). Using this approach, the parity of electronic states can be probed by measuring the XUV polarization-dependent photoemission yield (called LD-ARPES). The photoemission matrix element can be expressed as $\langle \psi_f|\hat{\mathcal{H}}^\prime|\psi_i\rangle$, where $\psi_i$ and $\psi_f$ denote the initial- and final-state wavefunctions, respectively. Within the velocity gauge, the light-matter interaction Hamiltonian is approximated as $\hat{\mathcal{H}}^\prime\simeq\mathbf{A}_{\rm{pr}} \cdot \hat{\mathbf{p}}$, with $\mathbf{A}_{\rm{pr}}$ the vector potential of the probe pulse and $\hat{\mathbf{p}}$ the electron momentum operator. Due to symmetry considerations, when both the light incidence and the photoelectron detection planes coincide with the crystal mirror plane, the photoemission yield to odd-parity final states vanishes~\cite {damascelli2003angle}. Under these conditions, only photoemission to even (under reflection about the scattering plane) final states is allowed, and thus the symmetry of $\hat{\mathcal{H}}^\prime$ directly determines photoemission from which initial states' parity is allowed. For a $p$-polarized probe pulse with the electric field lying in the scattering plane $\mathcal{M}_{y}$ [noted as AC ($p$-pol.) probe pulse], the light-matter interaction Hamiltonian $\hat{\mathcal{H}}^\prime$ is even, so only initial states with even parity contribute to the signal ($\langle \psi_f|\hat{\mathcal{H}}^\prime|\psi_i\rangle = \langle + | + | + \rangle \neq 0$), whereas those with odd parity are symmetry-forbidden 
($\langle + | + | - \rangle = 0$). Conversely, for an $s$-polarized probe pulse polarized along the ZZ direction [noted as ZZ ($s$-pol.) probe pulse], $\hat{\mathcal{H}}^\prime$ is odd, selectively coupling to  odd-parity initial states
($\langle + | - | - \rangle \neq 0$) while suppressing even-parity contributions  
($\langle + | - | + \rangle = 0$)~\cite{jung2020black}. In the following, to rationalize our measurements using these well-defined selection rules, we will focus on the experimental geometry in which both the plane of light incidence and the photoelectron detection plane ($\Gamma-\rm{Y}$ high symmetry direction) coincide with the crystal mirror plane ($\mathcal{M}_y$, AC direction). Complementary data acquired with the same plane of light incidence but with the photoelectron detection plane aligned along the $k_{\mathrm{ZZ}}$ direction are presented in Supplementary Figs.~S3-S5~\cite{SM}.

\subsection{Symmetry of electronic valence and conduction bands}

\begin{figure}
\centering
\includegraphics[width=8.6cm]{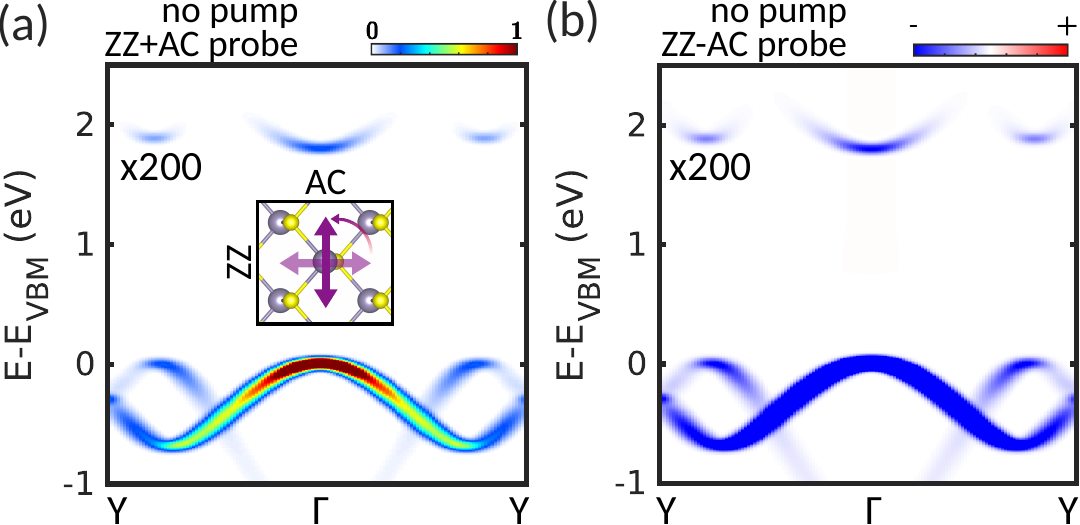}
\caption{\textbf{Valence and conduction band parity probed by TDDFT linear dichroism ARPES}. \textbf{(a)} The calculated valence and conduction bands ARPES intensity maps along the $k_{\mathrm{AC}}$ direction summed for both AC ($p$-pol.) and ZZ ($s$-pol.) probe pulses polarizations. The inset depicts the real-space in-plane structure of $n$-doped monolayer SnS with polarization-tunable XUV probe pulse (purple arrows). \textbf{(b)} The corresponding LD-ARPES highlighting the VB and CB parity probed through photoemission selection rules.}
\label{Fig2}
\end{figure}

First, to benchmark wavefunctions' parity and associated photoemission selection rules from VB and CB in equilibrium, we perform TDDFT simulations of ARPES intensity maps along the AC direction for $n$-doped monolayer SnS, using both AC ($p$-pol.) and ZZ ($s$-pol.) XUV probe fields [Fig.~\ref{Fig2}(a)]. In monolayer SnS, both the VB and CB wavefunctions along the $\Gamma-\rm{Y}$ direction are even under the mirror operation $M_{y}$. The light-matter interaction Hamiltonian $\hat{\mathcal{H}}^\prime$ transforms as odd for the ZZ ($s$-pol.) XUV probe and even for the AC ($p$-pol.) XUV probe under $M_{y}$. As a result, symmetry strictly forbids photoemission from both the VB and CB under the ZZ ($s$-pol.) XUV probe, whereas photoemission from these bands is symmetry-allowed under the AC ($p$-pol.) probe configuration (the symmetry analysis for bulk SnS yields identical results, see S1 and S2 of the SM~\cite{SM} for more details). The XUV LD-ARPES signal is defined as $I_{\rm s}-I_{\rm p}$, where $I_{\rm s}$ and $I_{\rm p}$ denote the photoemission intensities measured with the ZZ ($s$-pol.) and AC ($p$-pol.) XUV probe, respectively. Because the ZZ ($s$-pol.) configuration yields vanishing spectral weights for both the VB and CB, while the AC ($p$-pol.) probe produces a finite photoemission intensity, the LD signal is negative in the energy-momentum regions of these bands. This leads to the uniformly negative (blue) contrast of both the VB and CB observed in Fig.~\ref{Fig2}(b). Such a dichroic response provides a direct and symmetry-enforced fingerprint of the even parity of these states, fully consistent with the group-theory analysis.

\subsection{Floquet--Bloch states symmetry engineering}

\begin{figure}
\centering
\includegraphics[width=8.6cm]{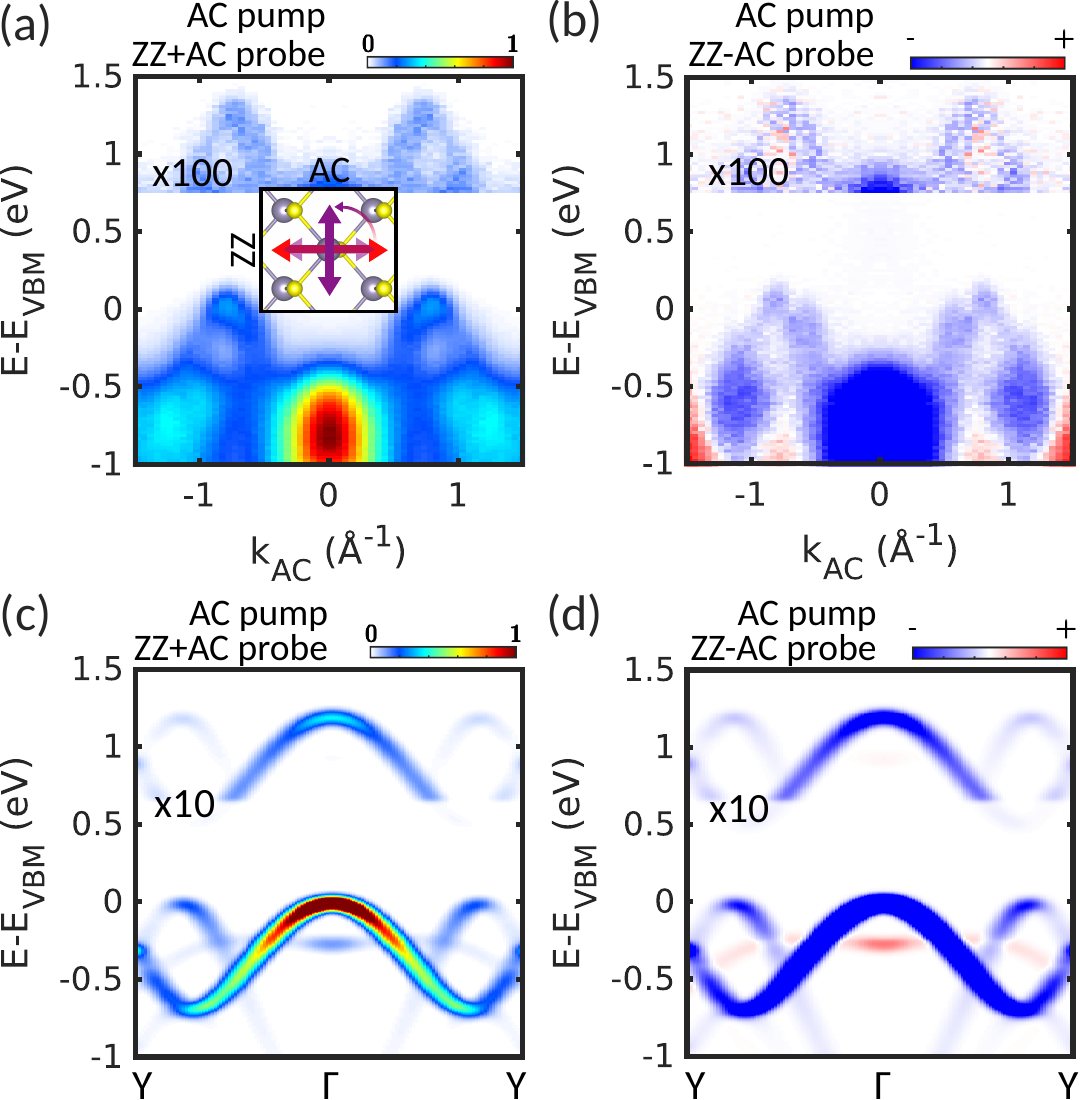}
\caption{\textbf{Emergence of Floquet--Volkov sidebands under AC ($p$-pol.) pump}. \textbf{(a)} Experimentally measured energy-momentum cuts along the $k_{\rm{AC}}$ direction at the pump-probe temporal overlap, using AC ($p$-pol.) pump and integrated for all XUV polarization angles, and \textbf{(b)} the associated LD-ARPES. The inset in (a) depicts the real-space in-plane structure of SnS with pump polarization along the AC direction (red arrow) and polarization-tunable probe (purple arrows). \textbf{(c)} Theoretically calculated energy-momentum cuts along the $k_{\rm{AC}}$ direction, using AC-polarized pump, obtained by TDDFT simulations by summing the spectra calculated for AC ($p$-pol.) and ZZ ($s$-pol.) probe pulses, and \textbf{(d)} the associated calculated LD-ARPES.}
\label{Fig3}
\end{figure}

Having established the photoemission selection rules governing the equilibrium valence and conduction bands, we now extend our study to the light-driven nonequilibrium regime, where the electronic structure is periodically dressed by a strong IR pump field. In contrast to Floquet states, which reflect the time-periodic dressing of Bloch electrons inside the crystal, Volkov states (or laser-assisted photoemission, LAPE) emerge from the dressing of outgoing photoelectrons by the pump field at the surface. The relative strength of these contributions, as well as their momentum-dependence, strongly depends on the polarization of the pump field~\cite{Mahmood16, Park14, merboldt24, choi24, ValleyFloquet24,Bao2025-zb}. Within this context, we investigate the emergence of pure Floquet--Bloch as well as Floquet--Volkov states in SnS, using different pump pulse polarizations. Figure~\ref{Fig3}(a) shows energy-momentum cuts along the $k_{\rm{AC}}$ direction at the pump-probe temporal overlap, measured using an AC ($p$-pol.) pump and integrated over all XUV polarization angles. While Floquet states are certainly present, in this experimental configuration, the Volkov contribution is expected to dominate, being maximized by the out-of-plane component of the incident pump field~\cite{keunecke_electromagnetic_2020, ValleyFloquet24}. As previously demonstrated in 2H-WSe$_2$~\cite{ValleyFloquet24} and owing to their microscopic origin, Volkov states are expected to exhibit the same XUV dichroism as the VB. The XUV LD-ARPES shown in Fig.~\ref{Fig3}(b) further supports this observation: the VB Y$^\prime$ valleys and their $+\hbar\omega$ sidebands display identical LD-ARPES signatures, with a clear preference for photoemission using AC ($p$-pol.) probe. Consequently, both the VB and the Volkov sidebands have even parity. Nevertheless, as discussed in detail below, symmetry analysis indicates that the AC ($p$-pol.) pump-induced Floquet states share the same parity as the VB, and coincidentally, the same parity as the Volkov states. 

The parity of Floquet--Bloch states is governed by the symmetries of both the material and the driving field. This can be naturally understood within the framework of Floquet optical selection rules~\cite{fan2025floquet}. In this formalism, the photoemission matrix element is also central, but the initial state is now the $n$-th light-induced sideband wavefunction of the VB at momentum $k$, denoted as $|\psi^{v,n}_{k}\rangle$. The symmetry of $\hat{\mathcal{H}}^\prime |\psi^{v,n}_k\rangle$ in the matrix element is jointly determined by the symmetry of the equilibrium VB wavefunction and by the pump-probe configuration. For bulk SnS at the $\mathrm{Y}^\prime$ point, AC ($p$-pol.) pumping combined with a ZZ ($s$-pol.) probe pulse leads to $\hat{\mathcal{H}}^\prime |\psi^{v,n}_{\mathrm{Y}^\prime}\rangle$ transforming as odd under the mirror operation $M_{y}$, which is independent of the Floquet index $n$ and suppresses the photoemission intensity. In contrast, when the probe polarization is switched to AC ($p$-pol.), $\hat{\mathcal{H}}^\prime |\psi^{v,n}_{\mathrm{Y}^\prime}\rangle$ becomes even under $M_{y}$ operation, allowing photoemission for all relevant sidebands, as summarized in Table~\ref{tab:trarpes} (actually, along the $\Gamma-\rm{Y}$ line, the symmetry of the $\hat{\mathcal{H}}^\prime |\psi_{\rm{k}}^{v,n}\rangle$, with respect to $\mathcal{M}_{y}$ is unchanged once we fixed the pump-probe configuration. See more details in S3 of SM~\cite{SM}). Consequently, the LD-ARPES signal exhibits negative intensity (blue contrast) for both the $n=0,1$ sidebands along the $\Gamma-\rm{Y}$ direction, as shown in Fig.~\ref{Fig3}(b). This feature is therefore a symmetry-enforced result rather than a band- or sideband-specific effect. 

We then extend the same symmetry analysis to monolayer SnS. Once the pump-probe configuration and the $\Gamma-\rm{Y}$ momentum path are fixed, the transformation properties of the relevant electronic states and light-matter interaction Hamiltonian are identical to those of the bulk system, as summarized in Table~\ref{tab:trarpes}. This establishes that monolayer SnS captures the essential symmetry features of the sidebands in bulk SnS. Motivated by this equivalence, we perform TDDFT simulations of the LD-ARPES for monolayer SnS. The calculated results, shown in Figs.~\ref{Fig3}(c) and~\ref{Fig3}(d), reproduce both the Floquet--Volkov sideband structure and the polarization-dependent intensity contrast observed experimentally [Figs.~\ref{Fig3}(a) and~\ref{Fig3}(b)]. The excellent agreement among experiment, symmetry analysis, and TDDFT simulations provides a consistent and unified understanding of the observed Floquet phenomena.

\begin{figure}
\centering
\includegraphics[width=8.6cm]{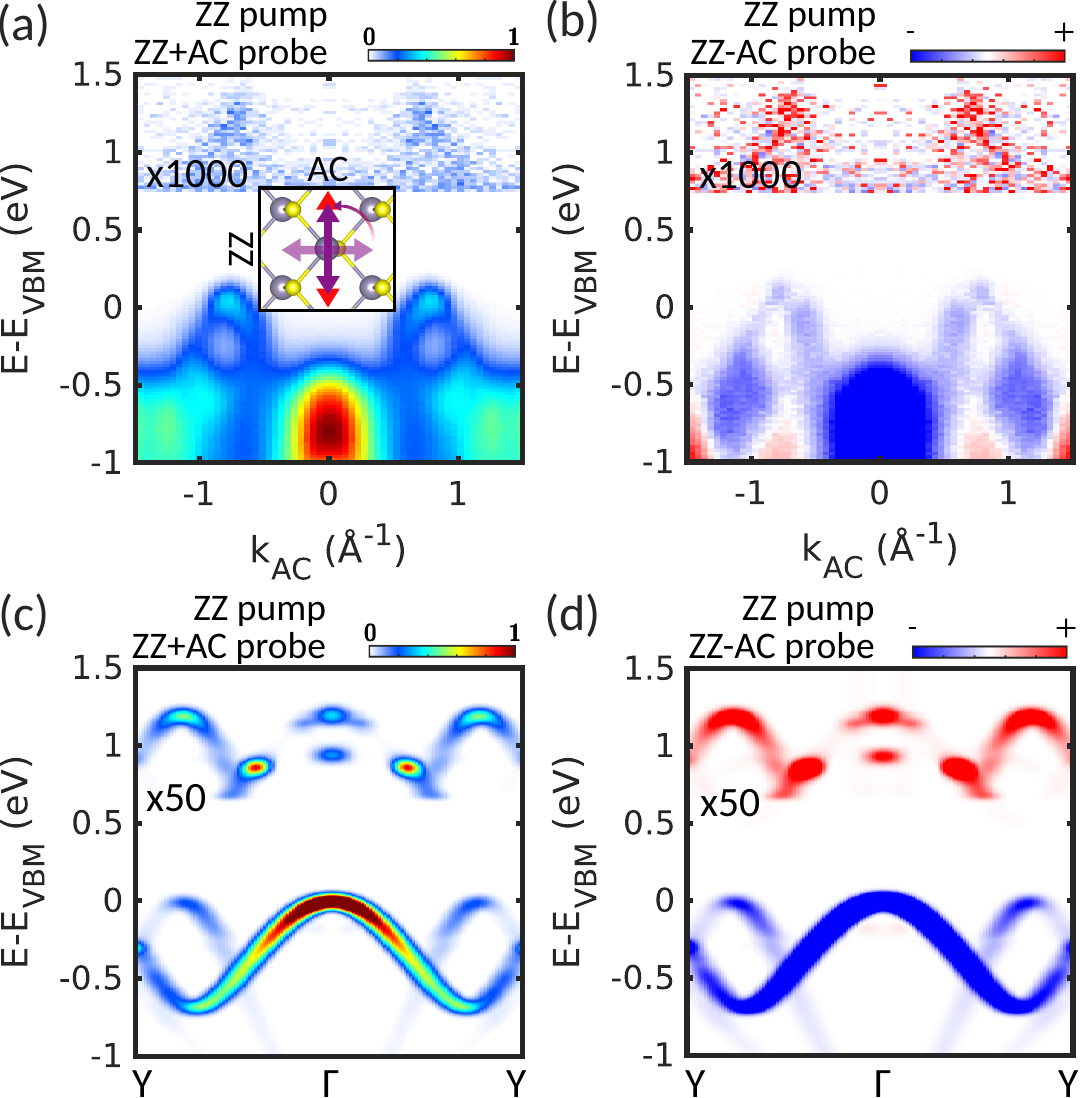}
\caption{\textbf{Parity switching of Floquet--Bloch states with ZZ ($s$-pol.) pump}. \textbf{(a)} Experimentally measured energy-momentum cuts along the $k_{\rm{AC}}$ direction at the pump-probe temporal overlap, using ZZ ($s$-pol.) pump and integrated for all XUV polarization angles, and \textbf{(b)} the associated LD-ARPES. The inset in (a) depicts the real-space in-plane structure of SnS with IR pump polarization along the ZZ direction (red arrow) and polarization-tunable XUV probe (purple arrows). \textbf{(c)} Theoretically calculated energy-momentum cuts along the $k_{\rm{AC}}$ direction, using ZZ ($s$-pol.) pump, obtained by TDDFT simulations by summing the spectra calculated for AC ($p$-pol.) and ZZ ($s$-pol.) probe pulses, and \textbf{(d)} the associated calculated LD-ARPES.}
\label{Fig4}
\end{figure}

We now rotate the pump polarization to the ZZ ($s$-pol.) configuration and probe the electronic structure along the $k_{\rm{AC}}$ direction at pump-probe temporal overlap. In this geometry, the surface-projected electric field is oriented perpendicular to the scattering plane, causing the Volkov sideband amplitude to vanish for all momenta along the $k_{\rm{AC}}$ direction~\cite{keunecke_electromagnetic_2020, merboldt24}. The resulting energy–momentum cuts, integrated over all XUV polarization angles, are shown in Fig.~\ref{Fig4}(a). Consequently, the observed $+\hbar\omega$ sideband along the $k_{\rm{AC}}$ is of purely Floquet--Bloch origin. The corresponding LD-ARPES map in Fig.~\ref{Fig4}(b) reveals a reversal of the dichroic signal in the $+\hbar\omega$ Y$^\prime$ valley sideband relative to the VB, indicating that photoemission from the VB$+\hbar\omega$ sideband is now favored under ZZ ($s$-pol.) XUV light pulses. Similar to the previous symmetry analysis, here the pump polarization is fixed to ZZ ($s$-pol.), which modulates the $n$-th Floquet sideband wavefunction for the VB $|\psi^{v,n}_{k}\rangle$. Under this configuration, the symmetry of $\hat{\mathcal{H}}^\prime|\psi^{v,n}_{k}\rangle$ with respect to the mirror operation $M_{y}$ depends explicitly on the Floquet index $n$. For a ZZ ($s$-pol.) probe pulse, $\hat{\mathcal{H}}^\prime|\psi^{v,n}_{\mathrm{Y}^\prime}\rangle$ is odd (even) for even (odd) $n$, while switching the probe polarization to AC ($p$-pol.) reverses these symmetry properties, as listed in Table~\ref{tab:trarpes}. The symmetry considerations applied to other $k$ points along the $\Gamma-\rm{Y}$ direction yield identical results (see S3 in SM~\cite{SM}). As a consequence, for the $n=0$ Floquet sideband, photoemission is symmetry-forbidden under the ZZ ($s$-pol.) probe but allowed under the AC ($p$-pol.) probe, yielding a negative LD-ARPES signal (blue contrast). In contrast, for the $n=1$ sideband, the situation is reversed: photoemission is allowed for the ZZ ($s$-pol.) probe and suppressed for the AC ($p$-pol.) probe, resulting in a positive LD-ARPES signal (red contrast), as shown in Fig.~\ref{Fig4}(b). Based on the identical symmetry analysis results for monolayer SnS with bulk SnS listed in Table~\ref{tab:trarpes}, we further perform TDDFT simulations of the LD-ARPES maps. The simulated results in Figs.~\ref{Fig4}(c) and~\ref{Fig4}(d) reproduce the observed sign reversal between the $n=0$ and $n=1$ Floquet sidebands, fully capturing the experimentally observed polarization-dependent Floquet dynamics.

Table \ref{tab:trarpes} summarizes the symmetry analysis of the photoemission matrix element at the $\mathrm{Y'}$ point, whose little group is $C_{2v}$ in both bulk and monolayer SnS. By constructing the irreducible representation of $\hat{\mathcal{H}}'|\psi^{v,n}_{\mathrm{Y'}}\rangle$ in the trARPES matrix element, one can determine its parity under the mirror operation $M_y$, which dictates whether the corresponding photoemission channel is symmetry allowed. The analysis shows that the parity of the light-induced sidebands depends on the VB wavefunction at the $\mathrm{Y'}$ point, the pump polarization and the Floquet index $n$: AC ($p$-pol.) pump preserves even parity, whereas ZZ ($s$-pol.) pump introduces a parity alternation with $n$, and mixed pump-probe geometries interchange the selection rules. This group-theory result provides a general symmetry framework for inferring even/odd parity selection rules. In particular, the parity of the light-driven state, combined with the probe polarization, determines whether photoemission is allowed.

\begin{table}
\small
\centering
\caption{Group symmetry analysis of $\hat{\mathcal{H}}^\prime|\psi_{\rm{Y}^\prime}^{v,n}\rangle$ as a part of the trARPES matrix element at the $\rm{Y}^\prime$ point for both bulk and monolayer SnS under four representative configurations.}
\begin{tabular}{|c|c|c|}
\hline
\multirow[c]{2}{*}{\rule{0pt}{4.8ex}\makecell{\textbf{SnS}\\pump@probe}}
 & \multicolumn{2}{c|}{$\rm{Y}^\prime$ point: $C_{2v}$} \\ \cline{2-3}

 & \begin{tabular}[c]{@{}c@{}}$\hat{\mathcal{H}}^\prime|\psi_{\rm{Y}^\prime}^{v,n}\rangle$ group\\
representation\end{tabular}
 & Symmetry \\
\hline

AC@ZZ
& $\Gamma_2\otimes(\Gamma_1\oplus\Gamma_4)^{|n|}\otimes\Gamma_1$
& odd \\
\hline

AC@AC
& $(\Gamma_1\oplus\Gamma_4)\otimes(\Gamma_1\oplus\Gamma_4)^{|n|}\otimes\Gamma_1$
& even \\
\hline

ZZ@ZZ
& $\Gamma_2\otimes(\Gamma_2)^{|n|}\otimes\Gamma_1$
& \begin{tabular}[c]{@{}c@{}}odd ($n$ is even)\\
even ($n$ is odd)\end{tabular} \\
\hline

ZZ@AC
& $(\Gamma_1\oplus\Gamma_4)\otimes(\Gamma_2)^{|n|}\otimes\Gamma_1$
& \begin{tabular}[c]{@{}c@{}}even ($n$ is even)\\
odd ($n$ is odd)\end{tabular} \\
\hline

\end{tabular}
\label{tab:trarpes}
\end{table}

\begin{figure}
\centering
\includegraphics[width=8.6cm]{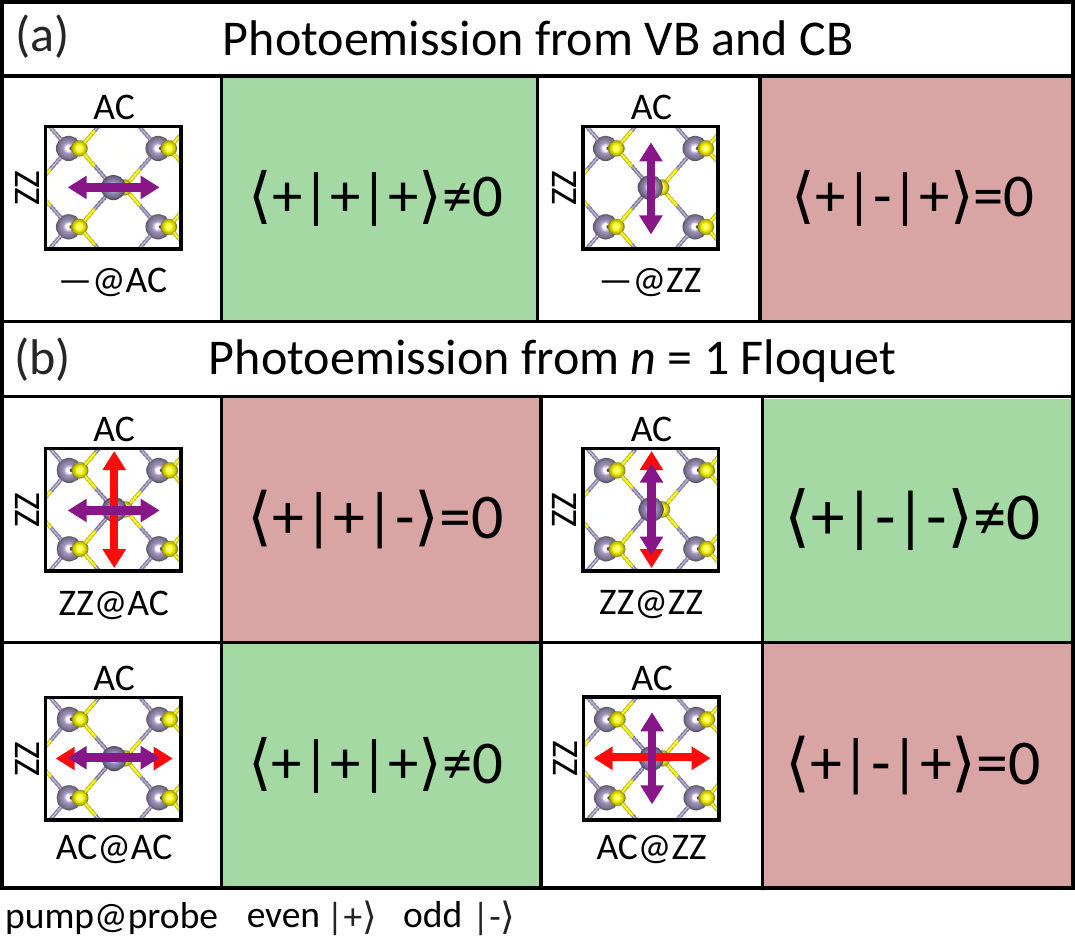}
\caption{\textbf{Photoemission selection rules in light-driven SnS}. \textbf{(a)} Symmetry-allowed (light green) and symmetry-forbidden (light red) photoemission matrix elements for the VB and CB at the Y$^\prime$ valleys under different XUV probe polarizations. 
\textbf{(b)} Symmetry-allowed (light green) and symmetry-forbidden photoemission matrix elements of the $n=1$ Floquet--Bloch states of the VB at the Y$^\prime$ valleys under different combinations of IR pump and XUV probe polarizations.}
\label{Fig5}
\end{figure}

A summary of the photoemission selection rules in light-driven SnS at the Y$^\prime$ valleys is presented in Fig.~\ref{Fig5} under polarization-tunable IR pump and XUV probe light pulses. Figure~\ref{Fig5}(a) illustrates the symmetry-dictated photoemission matrix elements of the even-parity VB and CB, which are allowed (or forbidden) under an XUV probe with even (or odd) parity. Additionally, Fig.~\ref{Fig5}(b) shows the symmetry-dictated photoemission matrix elements of the $n=1$ light-induced sideband for the VB. The parity of this light-driven state is determined by the polarization of the IR pump pulse and the VB, which in turn governs its relative polarization-dependent photoemission yield.

\subsection{Polarization- and valley-selective Floquet-induced band renormalization}

Under periodic electromagnetic dressing, Floquet–Bloch states can hybridize with equilibrium states. This hybridization, known as the optical Stark effect, typically results in energy repulsion between equilibrium and Floquet states~\cite{Autler1955-st,Frohlich1985-yn}. However, band renormalization via the optical Stark effect occurs only when the light-induced Floquet and neighboring equilibrium bands possess the appropriate relative symmetry to enable hybridization. Such symmetry-enforced selection rules for Floquet-induced band renormalization have been demonstrated in transition-metal dichalcogenides (TMDCs) through the valley-selective optical Stark effect~\cite{Sie15, Sie17}, where band hybridization is allowed only when the equilibrium and Floquet states share the same magnetic quantum number $m$~\cite{Sie15, Sie17}. More recently, Floquet-induced renormalization has also been shown to be pseudospin-dependent in semiconducting black phosphorus~\cite{Zhou23,Zhou23_2}. In the previous sections, we demonstrated the ability to control the parity of the Floquet-band wavefunctions in a polarization- and valley-selective manner in SnS. This makes SnS a compelling platform for investigating and generalizing the symmetry-enforced selection rules governing Floquet-induced band renormalization in multivalley systems with distinct crystal symmetries.

\begin{figure}
\centering
\includegraphics[width=8.6cm]{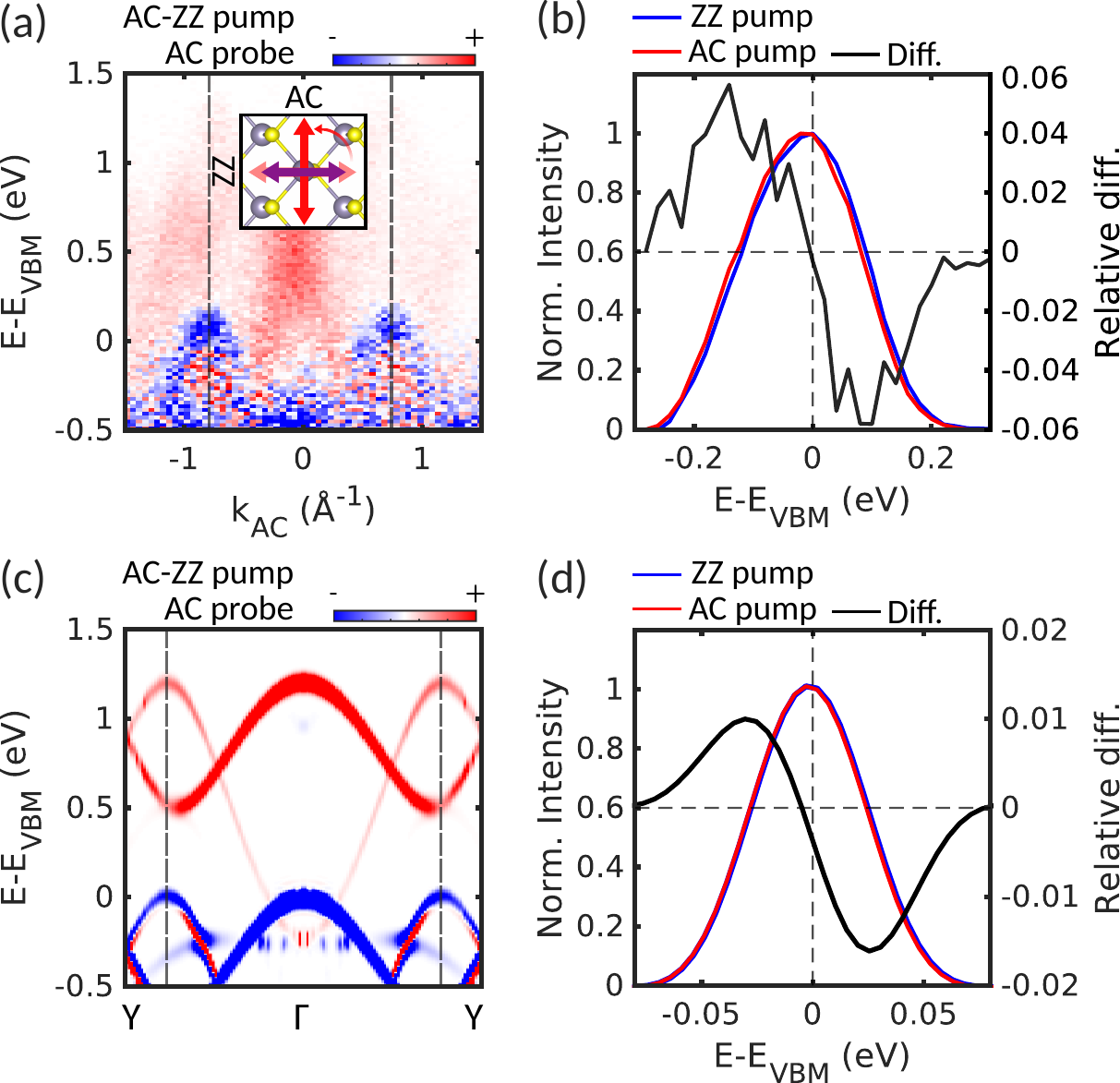}
\caption{\textbf{Light-induced Floquet band renormalization}. \textbf{(a)} Pump-polarization differential energy–momentum cut along the $k_{\mathrm{AC}}$ direction using AC ($p$-pol.) XUV probe pulses, measured at pump–probe temporal overlap. The signal is obtained by subtracting the photoemission spectra measured with AC ($p$-pol.) and ZZ ($s$-pol.) IR driving pulses, highlighting the emergence of light-induced Floquet band renormalization for AC ($p$-pol.) pump fields. Dashed markers indicate where the EDCs in (b) were extracted. The inset in (a) depicts the real-space in-plane structure of SnS with probe polarization along the AC direction (purple arrow) and polarization tunable pump (red arrows). \textbf{(b)} EDCs at the Y$^\prime$ valleys for AC ($p$-pol.) and ZZ ($s$-pol.) pump pulses, together with their relative difference. \textbf{(c)} Theoretical differential energy–momentum cut along the $k_{\mathrm{AC}}$ direction using AC ($p$-pol.) XUV probe pulses at pump-probe temporal overlap, obtained from TDDFT simulations by subtracting spectra calculated for AC ($p$-pol.) and ZZ ($s$-pol.) IR driving fields. Dashed markers indicate where the EDCs in (d) were extracted. \textbf{(d)} Corresponding theoretical EDCs at the Y$^\prime$ valleys for AC ($p$-pol.) and ZZ ($s$-pol.) pump fields, and their relative difference.}
\label{Fig6}
\end{figure}

Motivated by these considerations, we investigated polarization- and valley-selective Floquet-induced band renormalization in the multivalley SnS. In the pump-polarization differential map in Fig.~\ref{Fig6}(a), measured with an AC ($p$-pol.) probe, around the Y$^\prime$ valley band edge, a prominent negative (blue) differential signal is observed, followed by a positive (red) signal at higher binding energies. This differential map suggests a renormalization (energy downshift) of the VB under AC pumping. To confirm that these features in the differential energy–momentum map originate from band renormalization, we extracted energy distribution curves (EDCs) at the Y$^\prime$ valley for the two pump-polarization configurations [Fig.~\ref{Fig6}(b)]. The normalized EDCs reveal a small ($\sim$8~meV) yet clear energy downshift under AC ($p$-pol.) pumping. In addition, the black curve in Fig.~\ref{Fig6}(b) shows the differential EDC between ZZ ($s$-pol.) and AC ($p$-pol.) pumping. This antisymmetric differential signal, characterized by a single negative dip followed by a positive peak, is indicative of a rigid energy shift associated with band renormalization.

This pronounced polarization selectivity of the band renormalization is a consequence of light-induced Floquet states parity control and their subsequent interaction with neighboring bands~\cite{Zhou23,Zhou23_2}, whereby hybridization between same-parity equilibrium and Floquet states leads to energy-level repulsion. In particular, the VB hybridizes with the $n=-1$ Floquet CB replica of the same (even) parity, resulting in an energy downshift of the VB by $\sim$8~meV, as illustrated in the schematic of Fig.~\ref{Fig1}(b). Conversely, in the ZZ pumping ($p$-pol.) case, the CB $n=-1$ replica acquires odd parity, keeping it symmetry-forbidden from hybridizing with the VB of opposite parity and therefore preventing any renormalization. Furthermore, as shown in the Supplementary Fig.~S5~\cite{SM}, no band renormalization is detected along the $k_{\rm{ZZ}}$ direction. This momentum-dependent behavior clearly distinguishes the observed polarization- and valley-specific renormalization effect from pump-induced space-charge or surface photovoltage-induced band shift, which would manifest as a rigid, momentum-independent displacement of the entire photoemission spectra. 

These observed polarization-dependent, symmetry-enforced selection rules governing Floquet-induced band renormalization in multivalley SnS are also well captured by our TDDFT calculations. The theoretical pump-polarization-dependent differential map [Fig.~\ref{Fig6}(c)], obtained from TDDFT, along with the corresponding EDCs [Fig.~\ref{Fig6}(d)] at the Y$^\prime$ valley of monolayer SnS show that the renormalization emerges exclusively for AC ($p$-pol.) pump pulses. The associated antisymmetric positive–negative differential signal is in excellent qualitative agreement with the experimental results shown in Fig.~\ref{Fig6}(b). Although the magnitude of the effect is smaller in the theoretical calculations than in the experimental measurements, these results confirm the Floquet origin of the valence-band renormalization. Additionally, it should be noted, that although the $n=1$ sideband signal is strongly dominated by Volkov contributions for AC ($p$-pol.) pump pulses, the renormalization observed under this configuration confirms the presence of Floquet--Bloch states coexisting with Volkov states. Moreover, this effect is polarization- and valley-selective, in stark contrast to the pseudospin-selective renormalization observed in black phosphorus, which is restricted to the $\Gamma$ point~\cite{Zhou23}. These results represent a significant advance, demonstrating not only the ability to control the parity of Floquet states via symmetry-enforced selection rules, but also to monitor their influence on the selective and tunable renormalization of equilibrium bands in complex multivalley systems.

\section{Conclusions}

In this work, we investigated symmetry-driven Floquet engineering in bulk multivalley SnS, highlighting the role of combined light-matter symmetry as a fundamental organizing principle for nonequilibrium electronic states. Using time-, polarization-, and angle-resolved XUV photoemission spectroscopy, complemented by group-theory analysis and TDDFT, we examined the polarization- and valley-dependent parity and renormalization of both equilibrium and light-dressed bands in below-bandgap driven SnS.

We first showed that the equilibrium valence and conduction bands obey symmetry-driven photoemission selection rules dictated by the orientation of the probe field relative to the crystal mirror plane. These rules naturally extend to nonequilibrium Floquet--Bloch and Floquet--Volkov states at high-momentum valleys. By exploiting the interplay between crystal symmetry and pump/probe polarization, we demonstrated deterministic control and established a general methodology to probe the parity of light-dressed electronic states. We further extended symmetry-enforced selection rules to describe Floquet-induced band renormalization in multivalley systems, revealing polarization- and valley-dependent modifications of the electronic structure.

These results are enabled by the recent development of a fully polarization-tunable ultrafast XUV beamline combined with time-of-flight momentum microscopy, allowing measurements of linear dichroism from nonequilibrium states across the entire BZ~\cite{Fragkos2025-ob}. This capability permits the study of Floquet-state parity at high-momentum valleys, beyond recent similar experiments restricted to states near the BZ center~\cite{Bao2024}. Our experimental approach for controlling and probing wavefunction parity is therefore broadly applicable to any light-driven quantum materials.

Overall, our findings extend symmetry-based Floquet selection rules to high-momentum valleys in anisotropic materials, establishing wavefunction symmetry as a tunable degree of freedom for Floquet engineering in low-symmetry quantum solids. The ability to manipulate both band dispersion and wavefunction symmetry through tailored light-matter interactions opens new avenues for dynamic control of optical and electronic responses and provides a general framework for symmetry-driven control in nonequilibrium quantum systems. In particular, light-induced Floquet states can transiently modify the optical properties~\cite{Tiwari2025-bx} and associated selection rules of semiconductors by creating in-gap states with controllable parity, which would exist only during the presence of the time-periodic drive. Moreover, this approach provides a platform for exploring and unraveling novel and hidden nonequilibrium phases, where wavefunction engineering may induce topologically nontrivial states.

\section{Methods}

\subsection{Experimental setup}

The experiments are conducted using a polarization-controllable ultrafast XUV beamline~\cite{Comby22} integrated with a momentum microscope end station~\cite{tkach24, tkach24-2, Fragkos2025-ob}. The light source is a high-repetition-rate Yb fiber laser (166~kHz, 1030~nm central wavelength, 135~fs pulse duration, 50~W average power; Amplitude Laser Group). For the XUV probe branch based on high-order harmonic generation (HHG), a substantial fraction of the laser output is frequency-doubled in a BBO crystal, yielding a few watts of 515~nm radiation. After adjusting the beam size and shaping the spatial profile to form an annular profile, the green pulses are focused into an argon gas jet, where XUV harmonics are generated. Residual 515~nm light is removed from the XUV beam using successive spatial filtering stages with pinholes. The 9th harmonic of the 515~nm driver (21.6~eV) is selected through a combination of Sc/SiC multilayer XUV mirror reflections (NTTAT) and transmission through a 200~nm-thick Sn filter (Luxel). The infrared pump arm utilizes a small portion of the fundamental 1030~nm pulses. Pump and probe beams are recombined collinearly via a drilled mirror and focused onto the sample, producing typical spot sizes of approximately 70~$\mu$m~$\times$~140~$\mu$m for the IR pump and 45~$\mu$m~$\times$~35~$\mu$m for the XUV probe. Bulk $\mathrm{SnS}$ crystals (HQ Graphene) are mounted on a Flag-style sample holder (Ferrovac), with a cylindrical ceramic post (Umicore) attached to the sample surface using UHV-compatible conductive epoxy (Epo-tek). Cleavage is performed in situ by mechanically striking the ceramic post under ultrahigh-vacuum conditions (base pressure 1$\times$10$^{-10}$~mbar). The freshly cleaved samples are transferred to a motorized hexapod for precise alignment inside the main chamber (base pressure 2$\times$10$^{-10}$~mbar, room temperature). Time-resolved photoemission experiments are carried out with a custom time-of-flight momentum microscope featuring a versatile front-lens system with multiple operating modes (GST mbH)~\cite{tkach24, tkach24-2}. Data binning is performed using an open-source processing framework~\cite{Xian20, Xian19_2}, which converts raw single-event measurements into calibrated and binned multidimensional datasets. Further details of the experimental apparatus are provided in Ref.~\cite{Fragkos2025-ob}.

\subsection{First-principles calculation}

Our theoretical simulations were all performed using the real-space grid-based code Octopus~\cite{andrade2015real,tancogne2020octopus}. We first describe the density functional theory (DFT) computational details of the ground-state calculations for both bulk and monolayer SnS. The Kohn-Sham (KS) equations were solved self-consistently with a relative convergence of the density as $10^{-7}$, and a real-space grid spacing of 0.36 Bohr was adopted throughout. The Pseudodojo Perdew-Burke-Ernzerhof (PBE) pseudopotentials were employed for all elements. For the sampling in the BZ, a $\Gamma$-centered $10\times12\times4$ ($10\times12\times1$) k-point grid was used for bulk (monolayer) SnS. The optimized in-plane lattice parameters of bulk (monolayer) SnS were 4.421 (4.486) \AA{} and 3.989 (4.017) \AA{} along the $x$ and $y$ directions, respectively. Along the $z$ direction, the bulk structure has a lattice constant of 11.174 \AA{}, while the monolayer SnS was modeled using non-periodic boundary conditions with a real-space length of 110 Bohr, ensuring that the absorbing boundaries operate efficiently without spurious reflections in the subsequent photoemission simulations. Spin-orbit coupling (SOC) effects were neglected in the present work.

The photoemission spectra of monolayer SnS were computed using the time-dependent surface flux method in the semi-periodic system (t-SURFFP)~\cite{tao2012photo,de2017first} within the framework of real-time real-space TDDFT. This method enables direct evaluation of photoemission probabilities from crystal surfaces without explicitly constructing continuum scattering states. The system was modeled as a slab geometry, periodic in-plane ($xy$-plane) and finite along the $z$ direction. The electron dynamics were governed by the time-dependent Kohn-Sham (TDKS) equation formulated in the velocity gauge and the dipole approximation: 
\begin{equation}
\mathrm{i}\frac{\partial}{\partial t}\left|\Psi_{\rm{n\mathbf{k}}}(t)\right\rangle=\left[\frac{1}{2}\left(-\mathrm{i}\mathbf{\nabla}-\frac{\mathbf{A}(t)}{c}\right)^2+V_{\rm{KS}}[\rho](t)\right]\left|\Psi_{\rm{n\mathbf{k}}}(t)\right\rangle
\label{eq:td_ks_eq}
\end{equation}
where $\left|\Psi_{\rm{n\mathbf{k}}}(t)\right\rangle$ is the KS orbital with the band index $n$ and momentum $\mathbf{k}$,  $\mathbf{A}(t)$ is the total laser vector potential, $c$ is the speed of light (in atomic units) and $V_{\rm{KS}}[\rho](t)$ is the time-dependent KS potential, which depends on the time-dependent electronic density $\rho$. The time step for propagation was set to $\Delta t=1.69\times10^{-3}$ fs to ensure numerical convergence.

To model the photoemission experiments, the total vector potential $\mathbf{A}(t)$ in Eq.~\eqref{eq:td_ks_eq} is decomposed as $\mathbf{A}(t)=\mathbf{A}_{\rm{pu}}(t)+\mathbf{A}_{\rm{pr}}(t)$, where
$\mathbf{A}_{\rm{pu}}(t)$ and $\mathbf{A}_{\rm{pr}}(t)$ correspond to the pumping and probe pulse fields, respectively. The pumping pulse, responsible for inducing sidebands, is expressed as
\begin{equation}
\mathbf{A}_{\rm{pu}}(t)=\mathbf{A}_{\rm{pu}}\sin^2\left(\frac{\pi t}{T_{\rm{pu}}}\right)\cos(\omega t)\Theta(T_{\rm{pu}}-t)\Theta(t)
\label{eq:pump_laser}
\end{equation}
with $\mathbf{A}_{\rm{pu}}=\frac{cE_{\rm{pu}}}{\omega}\boldsymbol{\epsilon}_{\rm{pu}}$, where $E_{\rm{pu}}$ is the peak electric field, $\boldsymbol{\epsilon}_{\rm{pu}}$ is the polarization direction, $\omega$ is the frequency, and $T_{\rm{pu}}$ is the total pulse duration of the pumping laser. $\Theta(t)$ denotes the time-dependent Heaviside step function, which ensures that the pumping laser field $\mathbf{A}_{\rm{pu}}(t)$ is nonzero only within the time window $t\in[0,T_{\rm{pu}}]$. In our simulations, we set the pump duration to $T_{\rm{pu}}=135$~fs, use a laser intensity of $8.7\times10^9$~W/cm$^2$, and take a photon energy of $\hbar\omega=1.2$~eV, where $\hbar$ is the reduced Planck constant. These parameters are chosen to match the experimental conditions to ensure a faithful comparison.

Similarly, the probe pulse field $\mathbf{A}_{\rm{pr}}(t)$ is given by
\begin{equation}
\mathbf{A}_{\rm{pr}}(t)=\mathbf{A}_{\rm{pr}}\sin^2\left(\frac{\pi t}{T_{\rm{pr}}}\right)\cos(\Omega t)\Theta(T_{\rm{pr}}-t)\Theta(t)
\label{eq:probe_laser}
\end{equation}
where the parameters follow the same definitions as in Eq.~\eqref{eq:pump_laser}, and we set $T_{\rm{pr}}=100$~fs, the peak intensity of $2\times10^5$~W/cm$^2$, and $\hbar\Omega=21.6$~eV. The specific pulse shapes adopted here are chosen for convenience, and the conclusions of this work are independent of these functional forms.

To obtain the photoelectron distribution, we computed the flux of the single-particle current through a surface $S$ located 30 Bohr away from the slab, where complex absorbing potentials were applied to eliminate unphysical rescattering. At the relatively low probe photon energy used here, photoelectron propagation is more sensitive to the details of the absorbing boundary; although convergence with respect to the simulation box size was carefully checked, small residual boundary-related interference effects cannot be completely excluded. The Volkov states, analytical solution of the time-dependent Schr{\"o}dinger equation for a free electron in an external light field, were used to describe photoelectrons in the vacuum:
\begin{equation}
\label{eq:volkov_state}
\chi_{\mathbf{p}}(\mathbf{r}, t) = \frac{1}{\sqrt{2\pi D}} e^{\mathrm{i} \mathbf{p} \cdot \mathbf{r}} e^{\mathrm{i} \Phi(\mathbf{p}, t)}
\end{equation}
Here $D$ is the unit-cell area, $\mathbf{p}$ is the total photoelectron momentum, and the Volkov phase $\Phi(\mathbf{p}, t)$ is given by $\Phi(\mathbf{p}, t) = -\frac{1}{2}\int_0^t \mathrm{d}\tau \left[ \mathbf{p} - \frac{\mathbf{A}(\tau)}{c} \right]^2$. The time-dependent expansion coefficient of the KS orbital $\left|\Psi_{n\mathbf{k}} (t)\right\rangle$ in Eq.~\eqref{eq:td_ks_eq} onto the Volkov states in Eq.~\eqref{eq:volkov_state} was obtained by integrating the flux of the current projected onto plane waves over time:
\begin{equation}
b_n(\mathbf{p},t) = - \int_0^t \mathrm{d}\tau \oint_S \mathrm{d}\mathbf{s} \cdot \left\langle \chi_{\mathbf{p}}(\tau)\right|\hat{\mathbf{j}}(\tau)\left|\Psi_{\rm{n\mathbf{k}}}(t)\right\rangle
\label{eq:tSURFF-final}
\end{equation}
where $\hat{\mathbf{j}}(t) = -\mathrm{i} \nabla - \mathbf{A}(t)/c$ is the single-particle current operator. The flux integration was performed over the surface $S$ whose normal is parallel to the non-periodic direction. Equation~(\ref{eq:tSURFF-final}) enables efficient evaluation of the photoemission signal using only the time-evolved wavefunctions at the surface $S$, where scattering states are well approximated by Volkov waves, making this method particularly suited for slab geometries. Therefore, the momentum-resolved photoelectron probability $P(\mathbf{p})$ was then obtained from the modulus squared of the expansion coefficient $b_n(\mathbf{p},t)$ in Eq.~\eqref{eq:tSURFF-final} as
\begin{equation}
\label{eq:Volkov_coe}
P(\mathbf{p}) = \lim_{t\rightarrow\infty}\frac{2}{N} \sum_{n=1}^{N/2} \left| b_n(\mathbf{p}, t) \right|^2
\end{equation}
where $N$ denotes the number of emitted electrons. 

Artificial doping of monolayer SnS was also explored to visualize CB contributions in the ARPES intensity plot by adding 0.025 electronic charges per unit cell in the main text. We note that such slight doping has minimal effects on the band structures.

\section*{Acknowledgments}
We thank Yann Mairesse for insightful discussions. We thank Nikita Fedorov, Romain Delos, Pierre Hericourt, Rodrigue Bouillaud, Laurent Merzeau, and Frank Blais for technical assistance. We thank Baptiste Fabre for implementing and maintaining the data binning code. This work is supported by the Cluster of Excellence 'CUI: Advanced Imaging of Matter' of the Deutsche Forschungsgemeinschaft (DFG) - EXC 2056 - project ID 39071599 and the European Research Council (ERC-2024-SyG- 101167294 ; UnMySt), We acknowledge support from the Max Planck-New York City Center for Non-Equilibrium Quantum Phenomena. The Flatiron Institute is a division of the Simons Foundation. We acknowledge the financial support of the IdEx University of Bordeaux/Grand Research Program "GPR LIGHT". This work is part of the ULTRAFAST and TORNADO projects of PEPR LUMA and was supported by the French National Research Agency, as a part of the France 2030 program, under grants ANR-23-EXLU-0002 and ANR-23-EXLU-0004. We acknowledge support from ERC Starting Grant ERC-2022-STG No. 101076639, the Marie Skłodowska-Curie Doctoral Networks TIMES No. 101118915 and SPARKLE No. 101169225, Quantum Matter Bordeaux, AAP CNRS Tremplin, AAP SMR from Université de Bordeaux, from the Italian Ministry of University and Research (MUR) under the PRIN 2022738 Grant No. 2022PX279E 003, and from the Next Generation EU Partenariato Esteso NQSTI-Spoke 2 (THENCE-740PE00000023) and MUR D.M. 737/2021 “Materials Manipulation with Light”. S.F. acknowledges funding from the European Union’s Horizon Europe research and innovation programme under the Marie Skłodowska-Curie 2024 Postdoctoral Fellowship No. 101198277 (TopQMat). 

The views and opinions expressed herein do not necessarily reflect those of the European Commission. Neither the European Union nor the granting authority can be held responsible for them.

\section*{Author contributions}
S.B. conceived the idea. S.F. and S.B. planned and performed experiments. S.F. and S.B. analyzed the experimental data. D.D. and S.P. participated in maintaining the laser system. B.F. performed the theoretical simulations and group-theory analysis with support from U.D.G., H.H., and A.R.. B.F., S.F., and S.B. analyzed the theoretical data. S.F., B.F., U.D.G., H.H., A.R., and S.B. contributed to the discussion and interpretation of the results. S.F., B.F., and S.B. wrote the manuscript with input from other authors.

\section*{Data availability}
The data that support the findings of this article are publicly available on the Zenodo open repository~\cite{zenodo_SnS_2026}.

\clearpage

\onecolumngrid

\begin{center}
\Large \textbf{\textsc{Supplemental Material: \\ Symmetry-Driven Floquet Engineering in Multivalley SnS }}\\    
\end{center}

\newcounter{SMmark}
\makeatletter

\@addtoreset{equation}{SMmark}
\@addtoreset{figure}{SMmark}
\@addtoreset{table}{SMmark}
\@addtoreset{page}{SMmark}
\@addtoreset{section}{SMmark}

\setcounter{figure}{0}
\setcounter{table}{0}
\setcounter{equation}{0}
\setcounter{section}{0}

\renewcommand{\thesection}{S\arabic{section}}  
\renewcommand{\thetable}{S\arabic{table}}  
\renewcommand{\thefigure}{S\arabic{figure}}
\renewcommand{\theequation}{S\arabic{equation}}

\section{Symmetry analysis of bulk tin sulfide}

In this work, we choose the armchair (AC) direction as $x$ direction and the zigzag (ZZ) direction as $y$ direction of the bulk SnS. In the absence of the spin degree, the bulk SnS belongs to the nonsymmorphic space group $\mathcal{G}$ (\textit{Pnma}, No.~62), with the translation group $\mathcal{T}$ as its invariant subgroup. So we can obtain its factor group $\mathcal{G/T}$ as
\begin{equation}
\mathcal{G/T}=\{E\mathcal{T},\tilde{C}_{2x}\mathcal{T},\tilde{C}_{2y}\mathcal{T},\tilde{C}_{2z}\mathcal{T},I\mathcal{T},\tilde{M}_x\mathcal{T},\tilde{M}_y\mathcal{T},\tilde{M}_z\mathcal{T}\}
\end{equation}
with $E$ is the identity element, $\tilde{C}_{2x}=\{C_{2x}|(\frac{1}{2},0,\frac{1}{2})\}$, $\tilde{C}_{2y}=\{C_{2y}|(0,\frac{1}{2},0)\}$, $\tilde{C}_{2z}=\{C_{2z}|(\frac{1}{2},\frac{1}{2},\frac{1}{2})\}$, $I=-E$, $\tilde{M}_x=\{M_x|(\frac{1}{2},0,\frac{1}{2})\}$, $\tilde{M}_y=\{M_y|(0,\frac{1}{2},0)\}$, $\tilde{M}_z=\{M_z|(\frac{1}{2},\frac{1}{2},\frac{1}{2})\}$. Actually, the factor group $\mathcal{G/T}$ is isomorphic to the Abelian group $D_{2h}$, which is the point group of the space group $\mathcal{G}$. 

At the $\Gamma(0,0,0)$ point, the group of the wave vector $\mathcal{G}_\Gamma$ is isomorphic to the point group $D_{2h}$. Consequently, our symmetry analysis is based on $D_{2h}$, whose character table is provided in \cref{tab:bulk_SnS_D2h}. Since it's an Abelian group, all irreducible representations are one-dimensional, resulting in non-degenerate bands at the $\Gamma$ point. Specifically, the valence band (VB) and conduction band (CB) wavefunctions, denoted as $\left|\psi^v_\Gamma\right\rangle$ and $\left|\psi^c_\Gamma\right\rangle$, transform as the irreducible representations $\Gamma_4^-$ and $\Gamma_2^+$, respectively. This ordering is the opposite of that in monolayer black phosphorus (BP)~\cite{fan2025floquet}.

\begin{table}[h]
\large
\centering
\caption{\bf{The character table for the point group $D_{2h}$ at the $\Gamma$ point.}}
\begin{tabular}{l|cccccccccc}
\hline & $E$ & $C_{2 x}$ & $C_{2 y}$ & $C_{2 z}$ & $I$ & $M_{x}$ & $M_{y}$ & $M_{z}$ & Basis & Bands\\
\hline$\Gamma_1^{+}$ & 1 & 1 & 1 & 1 & 1 & 1 & 1 & 1 & $x^2$, $y^2$, $z^2$ &\\
$\Gamma_2^{+}$ & 1 & -1 & 1 & -1 & 1 & -1 & 1 & -1 & $R_y$, $xz$ & CB\\
$\Gamma_3^{+}$ & 1 & 1 & -1 & -1 & 1 & 1 & -1 & -1 & $R_x$, $yz$ &\\
$\Gamma_4^{+}$ & 1 & -1 & -1 & 1 & 1 & -1 & -1 & 1 & $R_z$, $xy$ &\\
\hline$\Gamma_1^{-}$ & 1 & 1 & 1 & 1 & -1 & -1 & -1 & -1 & $xyz$ &\\
$\Gamma_2^{-}$ & 1 & -1 & 1 & -1 & -1 & 1 & -1 & 1 & $y$ &\\
$\Gamma_3^{-}$ & 1 & 1 & -1 & -1 & -1 & -1 & 1 & 1 & $x$ &\\
$\Gamma_4^{-}$ & 1 & -1 & -1 & 1 & -1 & 1 & 1 & -1 & $z$ & VB\\
\hline
\end{tabular}
\label{tab:bulk_SnS_D2h}
\end{table}
%We can change $x\rightarrow z$ and $z\rightarrow x$ to obtain the corresponding table in the Bilbao. 

Along the $\Gamma-\rm{Y}$ line [excluding the $\Gamma$ and $\rm{Y}(\frac{1}{2},0,0)$ points], the group of the wave vector $\mathcal{G}_{\Gamma-\rm{Y}}$ is isomorphic to the point group $C_{2v}$, which is the subgroup of the point group $D_{2h}$. Specifically, we focus on the $\rm{Y^\prime}$ point, and the VB and CB wavefunctions $\left|\psi^v_{\rm{Y^\prime}}\right\rangle$ and $\left|\psi^c_{\rm{Y^\prime}}\right\rangle$ both transform as the irreducible representations $\Gamma_1$, as shown in the character table of $C_{2v}$ in \cref{tab:bulk_SnS_X_C2v}.

\begin{table}[h!]
\large
\centering
\caption{\textbf{The character table for the point group $C_{2v}$ at the $\rm{Y^\prime}$ point.}}
\label{tab:bulk_SnS_X_C2v}
\begin{tabular}{c|cccccc}
\hline
 & E & $C_{2x}$ & $M_y$ & $M_z$ & Basis & Bands\\
\hline
 $\Gamma_1$ & 1 & 1  & 1  & 1  & $x$ & VB, CB\\
\hline
$\Gamma_2$ & 1 & -1  & -1 & 1 & $y$ &\\
\hline
$\Gamma_3$ & 1 & 1  & -1 & -1 &  & \\
\hline
$\Gamma_4$ & 1 & -1  & 1 & -1 & $z$ & \\
\hline
\end{tabular}
\end{table}

Along the $\Gamma-\rm{X}$ line [excluding the $\Gamma$ and $\rm{X}(0,\frac{1}{2},0)$ points], similarly, the group of the wave vector $\mathcal{G}_{\Gamma-\rm{X}}$ is isomorphic to the point group $C_{2v}$. Specifically, we focus on the $\rm{X^\prime}$ point, and the VB and CB wavefunctions $\left|\psi^v_{\rm{X^\prime}}\right\rangle$ and $\left|\psi^c_{\rm{X^\prime}}\right\rangle$ both transform as the irreducible representations $\Gamma_4$, as shown in the character table of $C_{2v}$ in \cref{tab:bulk_SnS_Y_C2v}.

\begin{table}[h]
\large
\centering
\caption{\textbf{The character table for the point group $C_{2v}$ at the $\rm{X^\prime}$ point.}}
\label{tab:bulk_SnS_Y_C2v}
\begin{tabular}{c|cccccc}
\hline
 & E & $C_{2y}$ & $M_x$ & $M_z$ & Basis & Bands\\
\hline
 $\Gamma_1$ & 1 & 1  & 1  & 1  & $y$ & \\
\hline
$\Gamma_2$ & 1 & -1  & 1 & -1 & $z$ &\\
\hline
$\Gamma_3$ & 1 & 1  & -1 & -1 &  & \\
\hline
$\Gamma_4$ & 1 & -1  & -1 & 1 & $x$ & VB, CB\\
\hline
\end{tabular}
\end{table}

\begin{figure}[h]
\includegraphics[width=0.6\columnwidth]{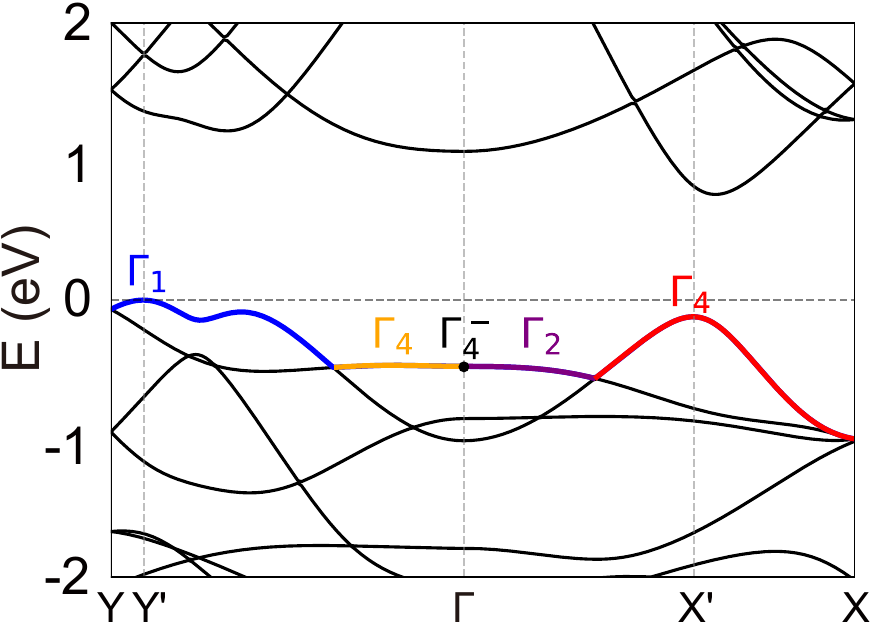}
\caption{Band structures of bulk SnS from the density functional theory (DFT) calculation. The irreducible representations of the VB are indicated. For points where the VB intersects with its neighboring band along the $\Gamma-\rm{Y}$ and $\Gamma-\rm{X}$ lines (excluding $\rm{Y}$ and $\rm{X}$ points with two-dimensional irreducible representations for the point group $D_{2h}$), the corresponding irreducible representations reduce to direct sums of the adjacent one-dimensional irreducible representations, since the point group $C_{2v}$ has only one-dimensional irreducible representations. These are omitted here for simplicity.}
\label{fig:bulk_SnS}
\end{figure}

Finally, we summarize the eigenvalues of the VB and CB wavefunctions at different $\bm{k}$ points for the bulk SnS respect to the mirror operation $M_y$, as shown in \cref{tab:bulk_SnS_eigenvalue}. Moreover, along the $\Gamma-\rm{Y}$ line, the VB wavefunctions are all even under the mirror operation $M_y$, with the VB and the corresponding irreducible representations highlighted in matching colors in \cref{fig:bulk_SnS}.

\begin{table}[h!]
\large
\centering
\caption{\textbf{The eigenvalues of different wavefunctions under the mirror operation $M_y$.}}
\label{tab:bulk_SnS_eigenvalue}
\begin{tabular}{c|cccc}
\hline
& $|\psi_{\rm{Y^\prime}}^{v,c}\rangle$ & $|\psi_{\Gamma}^{v}\rangle$ & $|\psi_{\Gamma}^{c}\rangle$ & $|\psi_{\rm{X^\prime}}^{v,c}\rangle$ \\
\hline
Irreducible Representation & $\Gamma_1$  & $\Gamma_4^-$  & $\Gamma_2^+$  & $\Gamma_4$\\
\hline
$M_y$ & 1  & 1 & 1 & --- \\
%\hline
%$M_x$ (ZZ scattering plane) & 0 & ---  & +1 & -1 & -1 & -2 & +2 \\
\hline
\end{tabular}
\end{table}

\clearpage
\section{Symmetry analysis of monolayer tin sulfide}

Similar to the bulk case, now we perform the symmetry analysis for monolayer SnS, taking the AC direction as $x$ direction and the ZZ direction as $y$ direction. In the absence of the spin degree, monolayer SnS belongs to the non-symmorphic space group \textit{Pmn2}$_1$ (No.~31), whose point group is $C_{2v}$. At the $\Gamma$ point, the VB and CB wavefunctions, $\left|\psi^v_\Gamma\right\rangle$ and $\left|\psi^c_\Gamma\right\rangle$, both transform as the irreducible representation $\Gamma_4$, as shown in the character table of $C_{2v}$ in \cref{tab:mono_SnS_Gamma_C2v}.

\begin{table}[h!]
\large
\centering
\caption{\textbf{The character table for the point group $C_{2v}$ at the $\Gamma$ point.}}
\label{tab:mono_SnS_Gamma_C2v}
\begin{tabular}{c|cccccc}
\hline
 & E & $C_{2x}$ & $M_z$ & $M_y$ & Basis & Bands\\
\hline
 $\Gamma_1$ & 1 & 1  & 1  & 1  & $x$ & \\
\hline
$\Gamma_2$ & 1 & -1  & 1 & -1 & $y$ &\\
\hline
$\Gamma_3$ & 1 & 1  & -1 & -1 &  & \\
\hline
$\Gamma_4$ & 1 & -1  & -1 & 1 & $z$ & VB, CB\\
\hline
\end{tabular}
\end{table}

% We can change $x\rightarrow z$, $y\rightarrow x$ and $z\rightarrow y$ to obtain the corresponding table in the Bilbao. 

Along the $\Gamma-\rm{Y}$ line, the group of the wave vector remains $C_{2v}$. At the representative point $\rm{Y^\prime}$, the VB and CB wavefunctions, $\left|\psi^v_{\rm{Y^\prime}}\right\rangle$ and $\left|\psi^c_{\rm{Y^\prime}}\right\rangle$, both transform as the irreducible representations $\Gamma_1$, as summarized in \cref{tab:mono_SnS_X_C2v}.

\begin{table}[h!]
\large
\centering
\caption{\textbf{The character table for the point group $C_{2v}$ at the $\rm{Y^\prime}$ point.}}
\label{tab:mono_SnS_X_C2v}
\begin{tabular}{c|cccccc}
\hline
 & E & $C_{2x}$ & $M_z$ & $M_y$ & Basis & Bands\\
\hline
 $\Gamma_1$ & 1 & 1  & 1  & 1  & $x$ & VB, CB\\
\hline
$\Gamma_2$ & 1 & -1  & 1 & -1 & $y$ &\\
\hline
$\Gamma_3$ & 1 & 1  & -1 & -1 &  & \\
\hline
$\Gamma_4$ & 1 & -1  & -1 & 1 & $z$ & \\
\hline
\end{tabular}
\end{table}

Along the $\Gamma-\rm{X}$ line (excluding the $\Gamma$ and \rm{X} points), the group of the wave vector reduces to $C_{s}$. At the representative point $\rm{X^\prime}$, the VB and CB wavefunctions, $\left|\psi^v_{\rm{X^\prime}}\right\rangle$ and $\left|\psi^c_{\rm{X^\prime}}\right\rangle$ both transform as the irreducible representation $\Gamma_1$, as shown in \cref{tab:mono_SnS_Cs}.

\begin{table}[h!]
\large
\centering
\caption{\textbf{The character table for the point group $C_{s}$ at the $\rm{X^\prime}$ point.}}
\label{tab:mono_SnS_Cs}
\begin{tabular}{c|cccc}
\hline
 & E & $M_z$ & Basis & Bands\\
\hline
 $\Gamma_1$ & 1 &1 & $x,y$ & VB, CB\\
\hline
$\Gamma_2$ & 1 & -1 & $z$ &\\
\hline
\end{tabular}
\end{table}

\begin{figure}[!htbp]
\includegraphics[width=0.6\columnwidth]{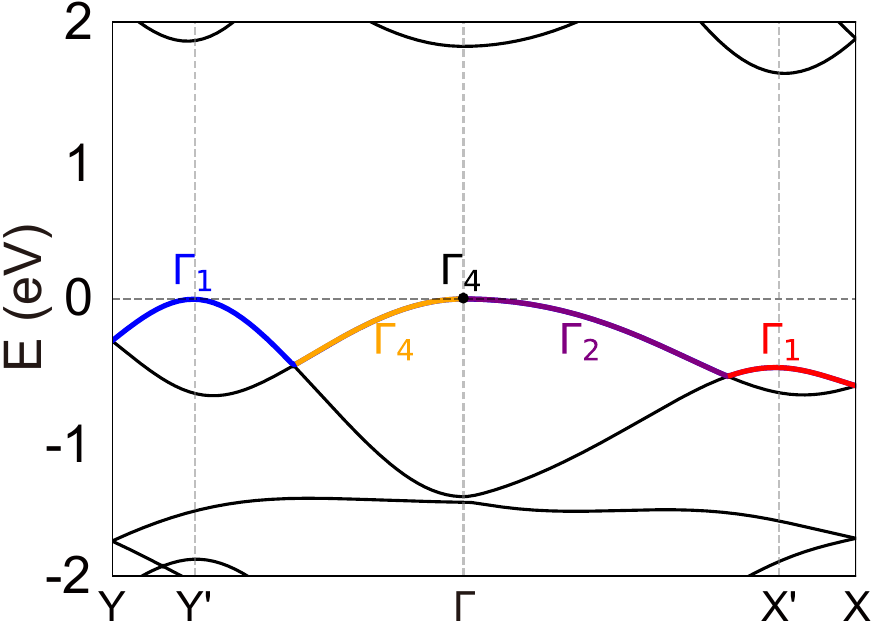}
\caption{Band structures of monolayer SnS from the DFT calculation, with the indirect gap as 1.63 eV. Similar to the bulk case, the irreducible representations of the VB are indicated.}
\label{fig:monolayer_SnS}
\end{figure}

Finally, we summarize the eigenvalues of the VB and CB wavefunctions at various $\bm{k}$ points with respect to the mirror operation $M_y$, as shown in \cref{tab:monolayer_SnS_eigenvalue}. Moreover, along the $\Gamma-\rm{Y}$ line, the VB wavefunctions are all even under the mirror operation $M_y$, and the VB and the corresponding irreducible representations are highlighted in matching colors in \cref{fig:monolayer_SnS}. This symmetry structure explains why monolayer SnS can fully capture the spectral features of bulk SnS for the AC scattering plane ($\mathcal{M}_y$).

\begin{table}[h!]
\large
\centering
\caption{\textbf{The eigenvalues of different wavefunctions under the mirror operation $M_y$.}}
\label{tab:monolayer_SnS_eigenvalue}
\begin{tabular}{c|ccc}
\hline
& $|\psi_{\rm{Y^\prime}}^{v,c}\rangle$ & $|\psi_{\Gamma}^{v,c}\rangle$ & $|\psi_{\rm{X^\prime}}^{v,c}\rangle$  \\
\hline
Irreducible Representation & $\Gamma_1$  & $\Gamma_4$  & $\Gamma_1$\\
\hline
$M_y$ & 1  & 1 & ---\\
%\hline
%$M_x$ (ZZ scattering plane) & 0 & ---  & +1 & -1 & -1 & -2 & +2 \\
\hline
\end{tabular}
\end{table}

%\clearpage

\section{Symmetry analysis of the photoemission spectroscopy}

The matrix element effect originates from Fermi's golden rule and depends on the symmetries of the initial and final states, as well as on the direction of the light polarization. Understanding this effect is crucial for interpreting photoemission spectra. Here we analyze the matrix element of bulk SnS using group theory.

Under the velocity gauge, the light-matter interaction Hamiltonian can be written as 
\begin{equation}
\hat{\mathcal{H}}^\prime\simeq\mathbf{A}_{pr}\cdot \mathbf{\hat{p}}    
\end{equation}
where $\mathbf{A}_{pr}$ is the vector potential of the external probe laser and $\mathbf{\hat{p}}$ is the electron momentum operator in bulk SnS~\cite{dresselhaus2007group}. The matrix element is then defined as
\begin{equation}
\langle\psi_{f}|\hat{\mathcal{H}}^\prime|\psi_{i}\rangle\simeq\mathbf{A}_{pr}\cdot\langle\psi_{f}|\mathbf{\hat{p}}|\psi_{i}\rangle
\end{equation}
with $\left|\psi_i\right\rangle$ and $\left|\psi_f\right\rangle$ denoting the wavefunctions of initial and final states, respectively. The wavefunction of the final state $\left|\psi_f\right\rangle$ should be \textbf{even} with respect to the scattering plane~\cite{damascelli2003angle}. Therefore, for nonzero photoemission intensity, the matrix element must be even under the mirror operation $M_{y}$. Since $\left|\psi_f\right\rangle$ is \textbf{even}, the direct product between the group representations of $\mathbf{\hat{p}}$ and $\left|\psi_i\right\rangle$ must also be \textbf{even} under $M_{y}$.

For clarity, two representative polarization configurations are defined:
\begin{itemize}
    \item ZZ ($s$-$pol.$): The probe laser is polarized purely along the ZZ ($y$) direction, without any out-of-plane ($z$) component.
    \item ZZ ($s$-$pol.$)@AC ($p$-$pol.$): The pumping laser is polarized along the ZZ ($y$) direction, while the probe laser lies in the scattering plane, containing AC ($x$) and $z$ components.
\end{itemize}

The following discussion focuses on how different configurations affect the matrix elements of ARPES and trARPES~\cite{fan2025floquet}.

\subsection{The matrix element of ARPES}
In this part we take the probe laser polarized along the ZZ ($s$-$pol.$) direction as an example to analyze the matrix element of ARPES. The vector potential $\mathbf{A}_{pr}$ of the probe laser is along the ZZ ($y$) direction, namely, $\hat{\mathcal{H}}^\prime\simeq A_{pr}^y\hat{p}_{y}$ transforms as the irreducible representation $\Gamma_2^-$ (\textbf{odd}):
\begin{enumerate}
\item If the initial state $\left|\psi_i\right\rangle$ is the VB wavefunction $\left|\psi^v_\Gamma\right\rangle$ at the $\Gamma$ point with the irreducible representation $\Gamma_4^-$, then we can obtain the irreducible representation of $\hat{\mathcal{H}}^\prime|\psi^v_\Gamma\rangle$ as
\begin{equation}
\underbrace{\Gamma_2^-}_{probe}\otimes\underbrace{\Gamma_4^-}_{\left|\psi^v\right\rangle}=\Gamma_3^+
\end{equation}
whose character under $M_y$ is -1 (\textbf{odd}). So around the $\Gamma$ point, the VB can't be detected under the ZZ ($s$-$pol.$) probe laser.

\item 
If the initial state $\left|\psi_i\right\rangle$ is the CB wavefunction $\left|\psi^c_\Gamma\right\rangle$ at the $\Gamma$ point with the irreducible representation $\Gamma_2^+$, in the same way we obtain the irreducible representation of $\hat{\mathcal{H}}^\prime|\psi^c_\Gamma\rangle$ as $\Gamma_2^-\otimes\Gamma_2^+=\Gamma_1^-$, whose character under $M_y$ is also -1 (\textbf{odd}). Similarly, around the $\Gamma$ point, the CB can also not be detected under the ZZ ($s$-$pol.$) probe laser.
\end{enumerate}

In summary, we list symmetry analyses of the matrix element at $\Gamma$ and $\rm{Y^\prime}$ points under different conditions for the probe laser in \cref{tab:arpes}. 

\begin{table}[h]
\normalsize
\centering
\caption{\bf{The symmetry analysis of the ARPES matrix element.}}
\begin{tabular}{|l|ll|ll|ll|}
\hline
\multirow{2}{*}{}    & \multicolumn{2}{c|}{Probe laser}       & \multicolumn{4}{c|}{Matrix element} \\ \cline{2-7}
& \multicolumn{1}{c|}{\begin{tabular}[c]{@{}c@{}}Polarization\\ direction\end{tabular}} & \multicolumn{1}{c|}{\begin{tabular}[c]{@{}c@{}}Group \\ representation\end{tabular}} & \multicolumn{1}{c|}{\begin{tabular}[c]{@{}c@{}}$\left|\psi_k^v\right\rangle$ group \\ representation\end{tabular}} & \multicolumn{1}{c|}{\begin{tabular}[c]{@{}c@{}}$\hat{\mathcal{H}}^\prime|\psi_k^v\rangle$ group\\
representation\end{tabular}} & \multicolumn{1}{c|}{\begin{tabular}[c]{@{}c@{}}$\left|\psi_k^c\right\rangle$ group \\ representation\end{tabular}} & \multicolumn{1}{c|}{\begin{tabular}[c]{@{}c@{}}$\hat{\mathcal{H}}^\prime|\psi_k^c\rangle$ group\\ representation\end{tabular}}  \\ 
\hline
\multicolumn{1}{|c|}{1. $\Gamma$ point} & \multicolumn{1}{c|}{ZZ ($s$-$pol.$)} &  \multicolumn{1}{c|}{$\Gamma_2^-$}     & \multicolumn{1}{c|}{$\Gamma_4^-$}   &  \multicolumn{1}{c|}{$\Gamma_3^+$, odd}   &    \multicolumn{1}{c|}{$\Gamma_2^+$} &
\multicolumn{1}{c|}{$\Gamma_1^-$, odd}\\ 
\hline
\multicolumn{1}{|c|}{2. $\Gamma$ point} & \multicolumn{1}{c|}{AC ($p$-$pol.$)} &  \multicolumn{1}{c|}{$\Gamma_3^-\oplus\Gamma_4^-$}     & \multicolumn{1}{c|}{$\Gamma_4^-$}   &  \multicolumn{1}{c|}{$\Gamma_2^+\oplus\Gamma_1^+$, even}   &    \multicolumn{1}{c|}{$\Gamma_2^+$} &
\multicolumn{1}{c|}{$\Gamma_4^-\oplus\Gamma_3^-$, even}\\ 
\hline
\multicolumn{1}{|c|}{3. $\rm{Y^\prime}$ point} & \multicolumn{1}{c|}{ZZ ($s$-$pol.$)} &  \multicolumn{1}{c|}{$\Gamma_2$}     & \multicolumn{1}{c|}{$\Gamma_1$}   &  \multicolumn{1}{c|}{$\Gamma_2$, odd}   &    \multicolumn{1}{c|}{$\Gamma_1$} &
\multicolumn{1}{c|}{$\Gamma_2$, odd}\\ 
\hline
\multicolumn{1}{|c|}{4. $\rm{Y^\prime}$ point} & \multicolumn{1}{c|}{AC ($p$-$pol.$)} &  \multicolumn{1}{c|}{$\Gamma_1\oplus\Gamma_4$}     & \multicolumn{1}{c|}{$\Gamma_1$}   &  \multicolumn{1}{c|}{$\Gamma_1\oplus\Gamma_4$, even}   &    \multicolumn{1}{c|}{$\Gamma_1$} &
\multicolumn{1}{c|}{$\Gamma_1\oplus\Gamma_4$, even}\\ 
\hline
\end{tabular}
\label{tab:arpes}
\end{table}

\subsection{The matrix element of trARPES}

Following the strategy in Ref.~\cite{fan2025floquet}, we now analyze the symmetry of the matrix element of trARPES. As a representative example, we focus on the $\Gamma$ point of the VB and consider both pumping and probe lasers polarized along the ZZ ($s$-$pol.$) direction with respect to the AC scattering plane.  

\paragraph{\textbf{Case 1}} --- ZZ ($s$-$pol.$)@ZZ ($s$-$pol.$):

The polarization directions of the pumping and probe laser are both totally along the ZZ direction. Suppose the wavefunction of the $n$-th sideband for the VB at the $\Gamma$ point is $|\psi^{v,n}_{\Gamma}\rangle$, we can obtain the representation of $\hat{\mathcal{H}}^\prime|\psi^{v,n}_{\Gamma}\rangle$ as
\begin{equation}
\underbrace{\Gamma_2^-}_{probe}\otimes(\underbrace{\Gamma_2^-}_{pump})^{|n|}\otimes\underbrace{\Gamma_4^-}_{\left|\psi^v_\Gamma\right\rangle}=(\Gamma_2^-)^{|n|+1}\otimes\Gamma_4^-
\end{equation}
\begin{enumerate}
\item If $n$ is \textbf{even}, then $(\Gamma_2^-)^{|n|+1}\otimes\Gamma_4^-=\Gamma_2^-\otimes\Gamma_4^-=\Gamma_3^+$, whose character under $M_y$ is -1 (\textbf{odd}). So around the $\Gamma$ point, the $n$-th sideband for the VB at the $\Gamma$ point can't be detected under the ZZ ($s$-$pol.$) probe laser.
\item 
If $n$ is \textbf{odd}, then $(\Gamma_2^-)^{|n|+1}\otimes\Gamma_4^-=\Gamma_1^+\otimes\Gamma_4^-=\Gamma_4^-$, whose character under $M_y$ is 1 (\textbf{even}). So around the $\Gamma$ point, the $n$-th sideband for the VB at the $\Gamma$ point can be detected under the ZZ ($s$-$pol.$) probe laser.
\end{enumerate}

\cref{tab:trarpes} summarizes the symmetry analyses for the matrix element at the $\Gamma$ and $\rm{Y^\prime}$ points under various pump-probe configurations. One can observe that $\hat{\mathcal{H}}^\prime|\psi^{v,n}_{\Gamma}\rangle$ and $\hat{\mathcal{H}}^\prime|\psi^{v,n}_{\rm{Y^\prime}}\rangle$ always share the same symmetry for identical pump-probe conditions. Examining the band structure along the $\Gamma-\rm{Y}$ line in \cref{fig:bulk_SnS}, the VB (blue) and the lower band intersect once. Along this direction, the symmetry is reduced from $D_{2h}$ at the $\Gamma$ point to $C_{2v}$. According to the compatibility relations in group theory, the $\Gamma_4^-$ irreducible representation at the $\Gamma$ point evolves into $\Gamma_4$ when the crystal momentum slightly departs from $\Gamma$ along the $\Gamma-\rm{Y}$ direction, which remains even under $M_y$ operation. Therefore, the symmetry of the VB (and its Floquet replica bands in trARPES) with respect to $M_y$ remains unchanged along the $\Gamma-\rm{Y}$ line. Moreover, we present a comparative symmetry analysis of monolayer and bulk SnS in \cref{tab:bulk_monolayer}, demonstrating that the monolayer SnS reliably captures the essential Floquet physics of the bulk system.

\begin{table}[h]
\normalsize
\centering
\caption{\bf{Symmetry analysis of the trARPES matrix element.}}
\begin{tabular}{|l|ll|ll|ll|}
\hline
\multirow{2}{*}{} &
\multicolumn{2}{c|}{Pumping laser} &
\multicolumn{2}{c|}{Probe laser} &
\multicolumn{2}{c|}{$\hat{\mathcal{H}}^\prime|\psi^{v,n}_k\rangle$, $k\in\{\Gamma, \rm{Y^\prime}\}$} \\ \cline{2-7}
& \multicolumn{1}{c|}{\begin{tabular}[c]{@{}c@{}}Polarization\\ direction\end{tabular}} 
& \multicolumn{1}{c|}{\begin{tabular}[c]{@{}c@{}}Group \\ representation\end{tabular}} 
& \multicolumn{1}{c|}{\begin{tabular}[c]{@{}c@{}}Polarization\\ direction\end{tabular}} 
& \multicolumn{1}{c|}{\begin{tabular}[c]{@{}c@{}}Group \\ representation\end{tabular}} 
& \multicolumn{1}{c|}{\begin{tabular}[c]{@{}c@{}}Group \\ representation\end{tabular}} 
& \multicolumn{1}{c|}{Symmetry} \\ 
\hline
\multicolumn{1}{|c|}{1. $\Gamma$ point} 
& \multicolumn{1}{c|}{AC ($p$-$pol.$)} & \multicolumn{1}{c|}{$\Gamma_3^-\oplus\Gamma_4^-$}     
& \multicolumn{1}{c|}{ZZ ($s$-$pol.$)} & \multicolumn{1}{c|}{$\Gamma_2^-$}   
& \multicolumn{1}{c|}{$\Gamma_2^-\otimes(\Gamma_3^-\oplus\Gamma_4^-)^{|n|}\otimes\Gamma_4^-$} 
& \multicolumn{1}{c|}{odd} \\ 
\hline
\multicolumn{1}{|c|}{2. $\Gamma$ point} 
& \multicolumn{1}{c|}{AC ($p$-$pol.$)} & \multicolumn{1}{c|}{$\Gamma_3^-\oplus\Gamma_4^-$}     
& \multicolumn{1}{c|}{AC ($p$-$pol.$)} & \multicolumn{1}{c|}{$\Gamma_3^-\oplus\Gamma_4^-$}   
& \multicolumn{1}{c|}{$(\Gamma_3^-\oplus\Gamma_4^-)\otimes(\Gamma_3^-\oplus\Gamma_4^-)^{|n|}\otimes\Gamma_4^-$} 
& \multicolumn{1}{c|}{even} \\ 
\hline
\multicolumn{1}{|c|}{3. $\Gamma$ point} 
& \multicolumn{1}{c|}{ZZ ($s$-$pol.$)} & \multicolumn{1}{c|}{$\Gamma_2^-$} 
& \multicolumn{1}{c|}{ZZ ($s$-$pol.$)} & \multicolumn{1}{c|}{$\Gamma_2^-$}
& \multicolumn{1}{c|}{$\Gamma_2^-\otimes(\Gamma_2^-)^{|n|}\otimes\Gamma_4^-$}
& \multicolumn{1}{c|}{\begin{tabular}[c]{@{}c@{}}odd ($n$ is even)\\ even ($n$ is odd)\end{tabular}} \\ 
\hline
\multicolumn{1}{|c|}{4. $\Gamma$ point} 
& \multicolumn{1}{c|}{ZZ ($s$-$pol.$)} & \multicolumn{1}{c|}{$\Gamma_2^-$}     
& \multicolumn{1}{c|}{AC ($p$-$pol.$)} & \multicolumn{1}{c|}{$\Gamma_3^-\oplus\Gamma_4^-$}   
& \multicolumn{1}{c|}{$(\Gamma_3^-\oplus\Gamma_4^-)\otimes(\Gamma_2^-)^{|n|}\otimes\Gamma_4^-$} 
& \multicolumn{1}{c|}{\begin{tabular}[c]{@{}c@{}}even ($n$ is even)\\ odd ($n$ is odd)\end{tabular}} \\ 
\hline
\multicolumn{1}{|c|}{5. $\rm{Y^\prime}$ point} 
& \multicolumn{1}{c|}{AC ($p$-$pol.$)} & \multicolumn{1}{c|}{$\Gamma_1\oplus\Gamma_4$}     
& \multicolumn{1}{c|}{ZZ ($s$-$pol.$)} & \multicolumn{1}{c|}{$\Gamma_2$}   
& \multicolumn{1}{c|}{$\Gamma_2\otimes(\Gamma_1\oplus\Gamma_4)^{|n|}\otimes\Gamma_1$} 
& \multicolumn{1}{c|}{odd} \\ 
\hline
\multicolumn{1}{|c|}{6. $\rm{Y^\prime}$ point} 
& \multicolumn{1}{c|}{AC ($p$-$pol.$)} & \multicolumn{1}{c|}{$\Gamma_1\oplus\Gamma_4$}     
& \multicolumn{1}{c|}{AC ($p$-$pol.$)} & \multicolumn{1}{c|}{$\Gamma_1\oplus\Gamma_4$}   
& \multicolumn{1}{c|}{$(\Gamma_1\oplus\Gamma_4)\otimes(\Gamma_1\oplus\Gamma_4)^{|n|}\otimes\Gamma_1$} 
& \multicolumn{1}{c|}{even} \\ 
\hline
\multicolumn{1}{|c|}{7. $\rm{Y^\prime}$ point} 
& \multicolumn{1}{c|}{ZZ ($s$-$pol.$)} & \multicolumn{1}{c|}{$\Gamma_2$} 
& \multicolumn{1}{c|}{ZZ ($s$-$pol.$)} & \multicolumn{1}{c|}{$\Gamma_2$}
& \multicolumn{1}{c|}{$\Gamma_2\otimes(\Gamma_2)^{|n|}\otimes\Gamma_1$}
& \multicolumn{1}{c|}{\begin{tabular}[c]{@{}c@{}}odd ($n$ is even)\\ even ($n$ is odd)\end{tabular}} \\ 
\hline
\multicolumn{1}{|c|}{8. $\rm{Y^\prime}$ point} 
& \multicolumn{1}{c|}{ZZ ($s$-$pol.$)} & \multicolumn{1}{c|}{$\Gamma_2$}     
& \multicolumn{1}{c|}{AC ($p$-$pol.$)} & \multicolumn{1}{c|}{$\Gamma_1\oplus\Gamma_4$}   
& \multicolumn{1}{c|}{$(\Gamma_1\oplus\Gamma_4)\otimes(\Gamma_2)^{|n|}\otimes\Gamma_1$} 
& \multicolumn{1}{c|}{\begin{tabular}[c]{@{}c@{}}even ($n$ is even)\\ odd ($n$ is odd)\end{tabular}} \\ 
\hline
\end{tabular}
\label{tab:trarpes}
\end{table}

\begin{table}
\small
\centering
\caption{\bf{The comparison of the trARPES matrix element at $\Gamma$ and $\rm{Y}^\prime$ points for the bulk and monolayer SnS under four representative configurations.}}
\begin{tabular}{|c|cc|cc|}
\hline
\multirow[c]{2}{*}{\rule{0pt}{4.8ex}\makecell{\textbf{Bulk SnS}\\pump@probe}}
 & \multicolumn{2}{c|}{$\Gamma$ point, $D_{2h}$ point group} & \multicolumn{2}{c|}{$\rm{Y}^\prime$ point, $C_{2v}$ point group} \\ \cline{2-5}
& \multicolumn{1}{c|}{\begin{tabular}[c]{@{}c@{}}$\hat{\mathcal{H}}^\prime|\psi_\Gamma^{v,n}\rangle$ group\\
representation\end{tabular}} & Symmetry & \multicolumn{1}{c|}{\begin{tabular}[c]{@{}c@{}}$\hat{\mathcal{H}}^\prime|\psi_{\rm{Y}^\prime}^{v,n}\rangle$ group\\
representation\end{tabular}} & Symmetry \\ 
\hline
\multicolumn{1}{|c|}{AC ($p$-pol.)@ZZ ($s$-pol.)} & \multicolumn{1}{c|}{$\Gamma_2^-\otimes(\Gamma_3^-\oplus\Gamma_4^-)^{|n|}\otimes\Gamma_4^-$} &  odd     & \multicolumn{1}{c|}{$\Gamma_2\otimes(\Gamma_1\oplus\Gamma_4)^{|n|}\otimes\Gamma_1$}   &  odd\\ 
\hline
\multicolumn{1}{|c|}{AC ($p$-pol.)@AC ($p$-pol.)} & \multicolumn{1}{c|}{$(\Gamma_3^-\oplus\Gamma_4^-)\otimes(\Gamma_3^-\oplus\Gamma_4^-)^{|n|}\otimes\Gamma_4^-$} &  even     & \multicolumn{1}{c|}{$(\Gamma_1\oplus\Gamma_4)\otimes(\Gamma_1\oplus\Gamma_4)^{|n|}\otimes\Gamma_1$}   &  even\\ 
\hline
\multicolumn{1}{|c|}{ZZ ($s$-pol.)@ZZ ($s$-pol.)} & \multicolumn{1}{c|}{$\Gamma_2^-\otimes(\Gamma_2^-)^{|n|}\otimes\Gamma_4^-$} &  \multicolumn{1}{c|}{\begin{tabular}[c]{@{}c@{}}odd ($n$ is even)\\
even ($n$ is odd)\end{tabular}}    & \multicolumn{1}{c|}{$\Gamma_2\otimes(\Gamma_2)^{|n|}\otimes\Gamma_1$}   &  \multicolumn{1}{c|}{\begin{tabular}[c]{@{}c@{}}odd ($n$ is even)\\
even ($n$ is odd)\end{tabular}}\\ 
\hline
\multicolumn{1}{|c|}{ZZ ($s$-pol.)@AC ($p$-pol.)} & \multicolumn{1}{c|}{$(\Gamma_3^-\oplus\Gamma_4^-)\otimes(\Gamma_2^-)^{|n|}\otimes\Gamma_4^-$} &  \multicolumn{1}{c|}{\begin{tabular}[c]{@{}c@{}}even ($n$ is even)\\ odd ($n$ is odd)\end{tabular}}   & \multicolumn{1}{c|}{$(\Gamma_1\oplus\Gamma_4)\otimes(\Gamma_2)^{|n|}\otimes\Gamma_1$}   &  \multicolumn{1}{c|}{\begin{tabular}[c]{@{}c@{}}even ($n$ is even)\\ odd ($n$ is odd)\end{tabular}} \\
\hline
\multirow[c]{2}{*}{\rule{0pt}{4.8ex}\makecell{\textbf{Monolayer SnS}\\pump@probe}}
 & \multicolumn{2}{c|}{$\Gamma$ point, $C_{2v}$ point group} & \multicolumn{2}{c|}{$\rm{Y}^\prime$ point, $C_{2v}$ point group} \\ \cline{2-5}
& \multicolumn{1}{c|}{\begin{tabular}[c]{@{}c@{}}$\hat{\mathcal{H}}^\prime|\psi_\Gamma^{v,n}\rangle$ group\\
representation\end{tabular}} & Symmetry & \multicolumn{1}{c|}{\begin{tabular}[c]{@{}c@{}}$\hat{\mathcal{H}}^\prime|\psi_{\rm{Y}^\prime}^{v,n}\rangle$ group\\
representation\end{tabular}} & Symmetry \\ 
\hline
\multicolumn{1}{|c|}{AC ($p$-pol.)@ZZ ($s$-pol.)} & \multicolumn{1}{c|}{$\Gamma_2\otimes(\Gamma_1\oplus\Gamma_4)^{|n|}\otimes\Gamma_4$} &  odd     & \multicolumn{1}{c|}{$\Gamma_2\otimes(\Gamma_1\oplus\Gamma_4)^{|n|}\otimes\Gamma_1$}   &  odd\\ 
\hline
\multicolumn{1}{|c|}{AC ($p$-pol.)@AC ($p$-pol.)} & \multicolumn{1}{c|}{$(\Gamma_1\oplus\Gamma_4)\otimes(\Gamma_1\oplus\Gamma_4)^{|n|}\otimes\Gamma_4$} &  even     & \multicolumn{1}{c|}{$(\Gamma_1\oplus\Gamma_4)\otimes(\Gamma_1\oplus\Gamma_4)^{|n|}\otimes\Gamma_1$}   &  even\\ 
\hline
\multicolumn{1}{|c|}{ZZ ($s$-pol.)@ZZ ($s$-pol.)} & \multicolumn{1}{c|}{$\Gamma_2\otimes(\Gamma_2)^{|n|}\otimes\Gamma_4$} &  \multicolumn{1}{c|}{\begin{tabular}[c]{@{}c@{}}odd ($n$ is even)\\
even ($n$ is odd)\end{tabular}}    & \multicolumn{1}{c|}{$\Gamma_2\otimes(\Gamma_2)^{|n|}\otimes\Gamma_1$}   &  \multicolumn{1}{c|}{\begin{tabular}[c]{@{}c@{}}odd ($n$ is even)\\
even ($n$ is odd)\end{tabular}}\\ 
\hline
\multicolumn{1}{|c|}{ZZ ($s$-pol.)@AC ($p$-pol.)} & \multicolumn{1}{c|}{$(\Gamma_1\oplus\Gamma_4)\otimes(\Gamma_2)^{|n|}\otimes\Gamma_4$} &  \multicolumn{1}{c|}{\begin{tabular}[c]{@{}c@{}}even ($n$ is even)\\ odd ($n$ is odd)\end{tabular}}   & \multicolumn{1}{c|}{$(\Gamma_1\oplus\Gamma_4)\otimes(\Gamma_2)^{|n|}\otimes\Gamma_1$}   &  \multicolumn{1}{c|}{\begin{tabular}[c]{@{}c@{}}even ($n$ is even)\\ odd ($n$ is odd)\end{tabular}} \\ 
\hline
\end{tabular}
\label{tab:bulk_monolayer}
\end{table}

% \clearpage

\section{Supporting experimental results}

\subsection{Floquet-Volkov states and linear dichroism (LD) along the $k_{\rm{ZZ}}$ direction}

Once the scattering plane is fixed to coincide with the mirror plane $\mathcal{M}_y$, the wavefunction along the $\Gamma-\rm{Y}$ line (AC direction) possesses a strictly well-defined parity. In contrast, along the $\Gamma–\rm{X}$ line (ZZ direction, excluding the $\Gamma$ and $\rm{X}$ points), the parity is not well defined under the mirror operation $M_y$. Nevertheless, an approximate odd parity can still be inferred for the X$^\prime$ valleys by examining the real-space electronic wavefunction at the X$^\prime$ point, as previously demonstrated by Tien et al.~\cite{Thanh_Tien2024-oo} on monolayer tin monochalcogenides. 

\begin{figure}[!htbp]
\centering
\includegraphics[width=0.6\columnwidth]{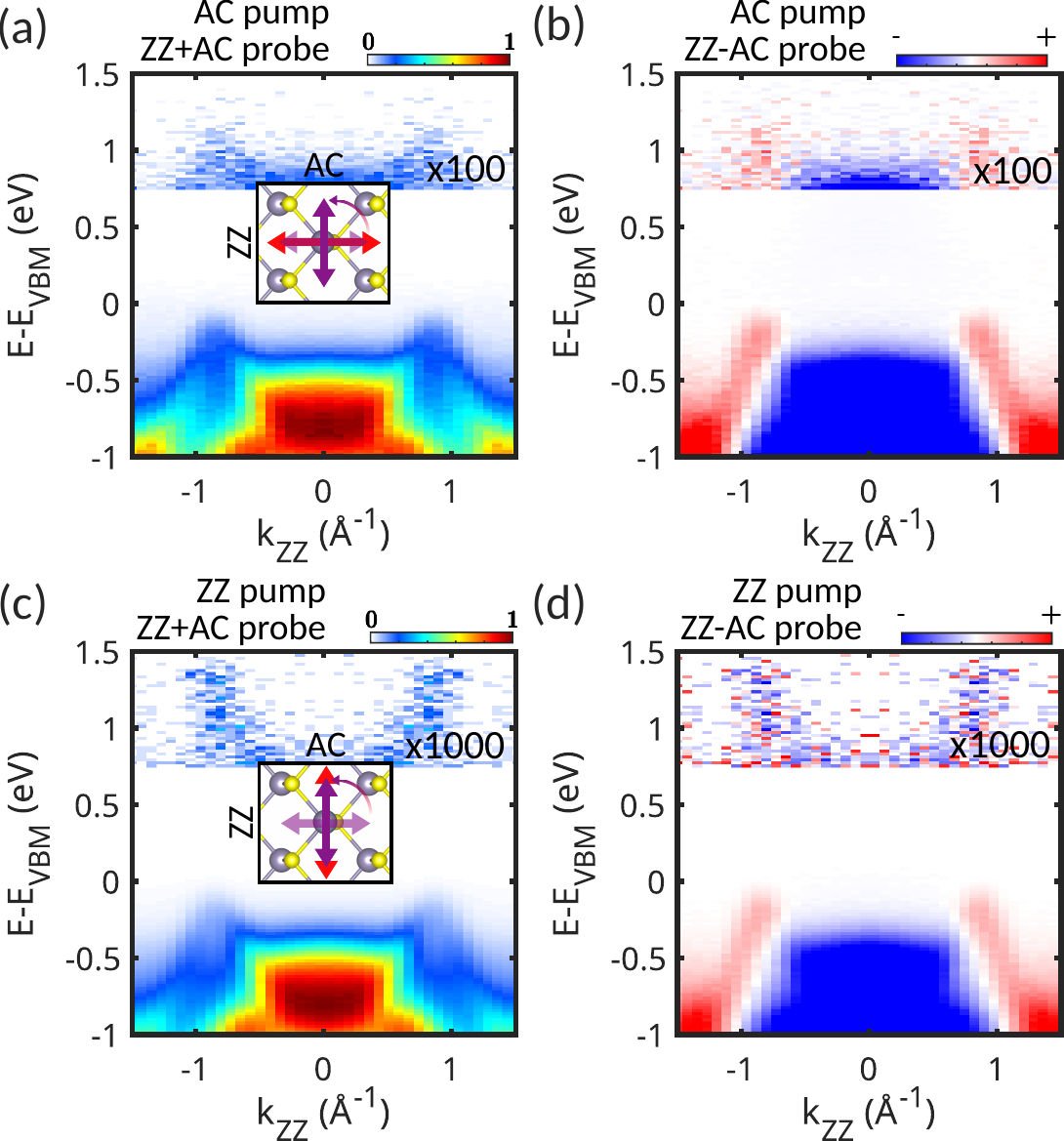}
\caption{\textbf{Emergence of Floquet--Volkov sidebands along the $k_{\rm{ZZ}}$ direction}. \textbf{(a)} Experimentally measured energy-momentum cuts along the $k_{\rm{ZZ}}$ direction at the pump-probe overlap, using AC ($p$-pol.) pump and integrated for all XUV polarization angles, and \textbf{(b)} the associated LD-ARPES. \textbf{(c)} Experimentally measured energy-momentum cuts along the $k_{\rm{ZZ}}$ direction at the pump-probe overlap, using ZZ ($s$-pol.) pump and integrated for all XUV polarization angles, and \textbf{(d)} the associated LD-ARPES.}
\label{kZZ_LDAD}
\end{figure}

We investigate the dichroic signal at the X$^\prime$ valley and for its $+\hbar\omega$ sideband. Figure~\ref{kZZ_LDAD}(a) presents energy-momentum cuts along the $k_{\mathrm{ZZ}}$ direction at pump-probe temporal overlap, obtained using an AC ($p$-pol.) pump and integrated over all extreme ultraviolet (XUV) polarization angles. Although Floquet states are certainly present, the Volkov contribution is expected to dominate under this experimental configuration, as already discussed in the main text. The XUV LD-ARPES map shown in Fig.~\ref{kZZ_LDAD}(b) indicates that the VB X$^\prime$ valleys and their $+\hbar\omega$ sidebands exhibit identical LD-ARPES signatures, with a clear preference for photoemission under a ZZ ($s$-pol.) XUV probe. Figure~\ref{kZZ_LDAD}(c) shows the experimental energy-momentum cuts along the $k_{\rm{ZZ}}$ direction at the pump-probe temporal overlap, obtained using a ZZ ($s$-pol.) pump and integrated over all XUV polarization angles. Clear sideband formation can be observed. Under this configuration, both Floquet-Bloch and Volkov states are expected to contribute. The corresponding LD-ARPES map in Fig.~\ref{kZZ_LDAD}(d) reveals a reversal of the dichroic signal in the $+\hbar\omega$ X$^\prime$ valley sideband relative to the VB, indicating that photoemission from the VB$+\hbar\omega$ sideband is now favored under AC ($p$-pol.) XUV probe. A similar reversal of the dichroic signal is also observed along the $k_{\mathrm{AC}}$ direction for ZZ ($s$-pol.) pump in Fig.~4 in the main text. As expected, the LD-ARPES signal at the X$^\prime$ valley is well described by our Floquet optical selection rules. Because the X$^\prime$ valley has approximately the opposite parity to the Y$^\prime$ valley, all selection-rule predictions under the same pump-probe configuration are systematically reversed.

\begin{figure}[!htbp]
\centering
\includegraphics[width=0.6\columnwidth]{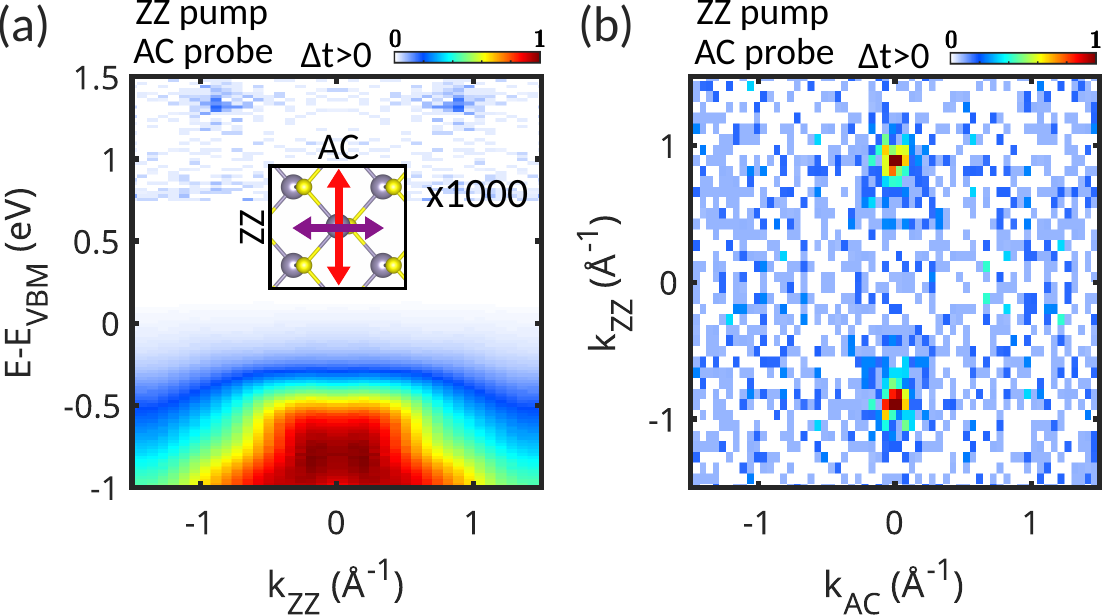}
\caption{\textbf{CB population at the X$^\prime$ valley}. \textbf{(a)} Energy-momentum cuts along the $k_{\mathrm{ZZ}}$ direction and \textbf{(b)} constants energy contour at $E-E_{\mathrm{VBM}}$=1.25~eV with $E_{\mathrm{VBM}}$ as the energy of the valence band maximum (VBM), obtained from a pump-probe scan (integration between 0.3 and 1 ps) with AC ($p$-pol.) probe and ZZ ($s$-pol.) pump, highlighting CB population.}
\label{CB}
\end{figure}

In contrast to the trARPES spectra obtained with an AC ($p$-pol.) pump in Fig.~\ref{kZZ_LDAD}(a,b), the spectra acquired with a ZZ ($s$-pol.) pump in Fig.~\ref{kZZ_LDAD}(c,d) exhibit nonvanishing photoemission intensity at the X$^\prime$ valleys at energies exceeding VB$+\hbar\omega$. This observation suggests some finite electron population at the conduction band minimum (CBM) at the X$^\prime$ valleys, even though the pump photon energy (1.2~eV) is slightly below the resonant direct interband transition at these high symmetry points. To clarify this behavior, we performed pump-probe measurements using an AC ($p$-pol.) XUV probe and a ZZ ($s$-pol.) pump. Figure~\ref{CB}(a) presents the experimental energy-momentum cut along the $k_{\mathrm{ZZ}}$ direction, integrated over positive pump-probe delays (integration between 0.3 and 1 ps). At excess energies $\sim$1.3 eV, the CBM at the X$^\prime$ valleys becomes visible and is further highlighted in the constant-energy contour at this excess energy [Fig.~\ref{CB}(b)]. Because these pockets lie more than 1.2~eV above the band edge at the X$^\prime$ valleys, they can only be populated via the high-energy tail of the pump pulse or through a two-photon absorption process.

\subsection{Polarization- and valley-selective renormalization}

\begin{figure}[!htbp]
\centering
\includegraphics[width=18cm]{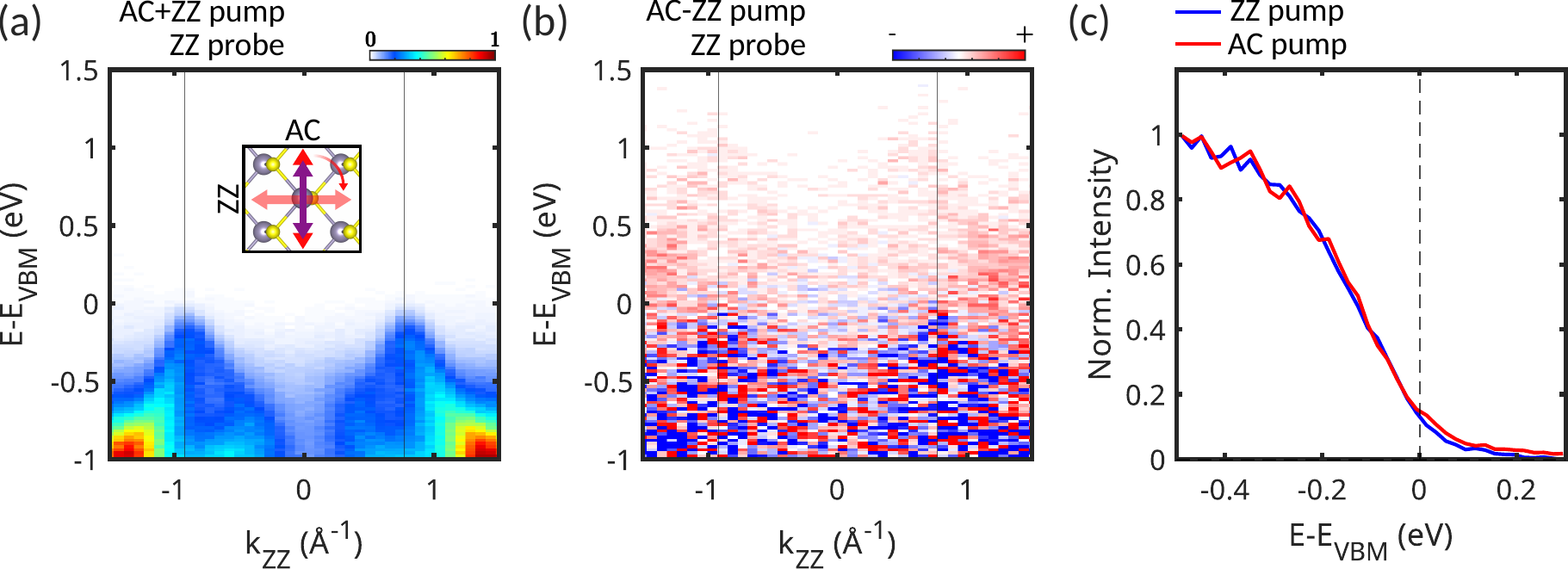}
\caption{\textbf{Absence of light-induced Floquet band renormalization along the $k_{\rm{ZZ}}$ direction}. \textbf{(a)} Energy-momentum cuts along the $k_{\rm{ZZ}}$ direction at the pump-probe overlap, using ZZ ($s$-pol.) XUV probe and integrated for all pump IR polarization angles. \textbf{(b)} Pump-polarization differential energy-momentum cut along the $k_{\rm{ZZ}}$ direction using ZZ ($s$-pol.) XUV probe pulses, measured at pump-probe overlap. The signal is obtained by subtracting the photoemission spectra measured with AC ($p$-pol.) and ZZ ($s$-pol.) IR pump driving pulses. Dashed markers in (a) and (b) indicate where the energy distribution curves (EDCs) in (c) were extracted. \textbf{(c)} EDCs at the X$^\prime$ valleys for AC ($p$-pol.) and ZZ ($s$-pol.) pump pulses.}
\label{renormal}
\end{figure}

As established in the main text, Floquet band renormalization along the $k_{\mathrm{AC}}$ direction is observed exclusively for AC ($p$-pol.) pump pulses. This pronounced polarization selectivity of the band renormalization is a consequence of light-induced Floquet states parity control and their subsequent interaction with neighboring bands, whereby hybridization between same-parity equilibrium and Floquet states leads to energy-level repulsion. In contrast, as shown in Fig.~\ref{renormal}, no clear signature of band renormalization is detected along the $k_{\mathrm{ZZ}}$ direction. At the X$^\prime$ valley the VB and CB are commonly assigned odd and even parity, respectively~\cite{Thanh_Tien2024-oo}. Within this symmetry picture, under ZZ ($s$-pol.) pump, the $n=-1$ Floquet replica of the CB would acquire odd parity and thus, in principle, allow hybridization with the VB, potentially leading to band renormalization. By contrast, for AC ($p$-pol.) pump, the $n=-1$ replica would retain even parity and remain symmetry-forbidden from hybridizing with the VB. However, the applicability of such parity-based selection rules is less clear at X$^\prime$, where the parity under the mirror operation $M_y$ is ill-defined, making symmetry-based expectations less definitive and complicating the assignment of the expected behavior. Moreover, Floquet-induced band renormalization is known to depend sensitively on detuning between the pump photon energy and the band gap and may exhibit non-monotonic behavior~\cite{Zhou23_2}. The specific detuning conditions in our experiment may therefore suppress the expected renormalization along $k_{\rm{ZZ}}$ under ZZ ($s$-pol.) pump. Importantly, the strong momentum dependence of the observed renormalization rules out trivial origins such as space-charge effects or surface photovoltage, which would instead result in a rigid, momentum-independent shift of the entire spectrum arising from modifications of the surface potential. This pronounced momentum selectivity therefore clearly establishes the light-induced nature of the observed Floquet band renormalization. 

%\bibliography{Main_SnS.bib}

\providecommand{\noopsort}[1]{}\providecommand{\singleletter}[1]{#1}%\providecommand{\noopsort}[1]{}\providecommand{\singleletter}[1]{#1}%

\end{document}